\pdfoutput=1
\documentclass[fleqn, usenatbib, usedcolumn]{mnras}

\newif\iflocal
\localfalse

\usepackage{graphicx}
\usepackage{xspace}
\usepackage{amsmath}
\usepackage{framed} 
\usepackage{txfonts}
\usepackage{epstopdf}
\usepackage{color}
\usepackage{rotating}
\usepackage{natbib}
\usepackage{ulem}
\usepackage{xspace}
\usepackage{sistyle}

\iflocal
\def\includedir{/Users/benedito/University/docs/latex}
\def\figdir{figs}
\else
\def\includedir{.}
\def\figdir{.}
\fi

\newcommand{\eqmn}[1]{equation~(\ref{#1})\xspace}

\def\gtsima{$\; \buildrel > \over \sim \;$}
\def\ltsima{$\; \buildrel < \over \sim \;$}
\def\prosima{$\; \buildrel \propto \over \sim \;$}
\def\gsim{\lower.7ex\hbox{\gtsima}}
\def\lsim{\lower.7ex\hbox{\ltsima}}
\def\simgt{\lower.7ex\hbox{\gtsima}}
\def\simlt{\lower.7ex\hbox{\ltsima}}
\def\simpr{\lower.7ex\hbox{\prosima}}

\newcommand{\dnorminl}[2]{{\rm d} #1 / {\rm d} #2}

\newcommand{\ndof}{N_{\rm dof}}
\newcommand\chidof{\chi^2/\ndof}

\newcommand{\vect}[1]{{\pmb #1}}

\newcommand{\gyr}{{\rm Gyr}}

\newcommand{\LCDM}{$\Lambda$CDM\xspace}

\newcommand{\rhom}{\rho_{\rm m}}

\def\sparta{\textsc{Sparta}\xspace}

\def\colossus{\textsc{Colossus}\xspace}

\def\rockstar{\textsc{Rockstar}\xspace}
\def\consistenttrees{\textsc{Consistent-Trees}\xspace}

\def\planck{Planck\xspace}
\def\wmap{WMAP7\xspace}

\def\erebos{Erebos\xspace}

\def\rmi{{\rm i}}

\def\rms{{\rm s}}
\def\rmt{{\rm t}}

\def\deltac{\delta_{\rm c}}
\def\rs{r_{\rm s}}

\def\gammadyn{\Gamma_{\rm dyn}}

\def\neff{n_{\rm eff}}

\def\rhoorb{\rho_{\rm orb}}

\def\ahalf{a_{1/2}}
\def\amerge{a_{\rm mm}}

\def\mtom{M_{\rm 200m}}
\def\rtom{R_{\rm 200m}}
\def\ctom{c_{\rm 200m}}

\def\nutom{\nu_{\rm 200m}}

\def\nutoc{\nu_{\rm 200c}}

\def\mtombnd{M_{\rm 200m,bnd}}
\def\mtomall{M_{\rm 200m,all}}

\def\rsp{R_{\rm sp}}

\newcommand{\rhos}{\rho_\rms}
\newcommand{\rt}{r_\rmt}
\newcommand{\rsteep}{r_{\rm steep}}

\newcommand{\rrs}{\left( \frac{r}{\rs} \right)}
\newcommand{\rrt}{\left( \frac{r}{\rt} \right)}
\newcommand{\rsrt}{\left( \frac{\rs}{\rt} \right)}

\newcommand{\rrsa}{\rrs^\alpha}
\newcommand{\rrse}{\rrs^\eta}
\newcommand{\rrtb}{\rrt^\beta}

\newcommand{\rsrtb}{\rsrt^\beta}

\newcommand{\delone}{\delta_{\rm 1}}
\newcommand{\delmax}{\delta_{\rm max}}

\usepackage{etoolbox}
\makeatletter
\patchcmd{\NAT@citex}
  {\@citea\NAT@hyper@{\NAT@nmfmt{\NAT@nm}\NAT@date}}
  {\@citea\NAT@nmfmt{\NAT@nm}\NAT@hyper@{\NAT@date}}
  {}
  {}
\patchcmd{\NAT@citex}
  {\@citea\NAT@hyper@{%
     \NAT@nmfmt{\NAT@nm}%
     \hyper@natlinkbreak{\NAT@aysep\NAT@spacechar}{\@citeb\@extra@b@citeb}%
     \NAT@date}}
  {\@citea\NAT@nmfmt{\NAT@nm}%
   \NAT@aysep\NAT@spacechar%
   \NAT@hyper@{\NAT@date}}
  {}
  {}
\patchcmd{\NAT@citex}
  {\@citea\NAT@hyper@{%
     \NAT@nmfmt{\NAT@nm}%
     \hyper@natlinkbreak{\NAT@spacechar\NAT@@open\if*#1*\else#1\NAT@spacechar\fi}%
       {\@citeb\@extra@b@citeb}%
     \NAT@date}}
  {\@citea\NAT@nmfmt{\NAT@nm}%
   \NAT@spacechar\NAT@@open\if*#1*\else#1\NAT@spacechar\fi%
   \NAT@hyper@{\NAT@date}}
  {}
  {}
\makeatother

\iflocal
\def\figdir{figs}
\else
\def\figdir{.}
\fi

\defcitealias{diemer_14}{DK14}
\defcitealias{diemer_22_prof1}{I}
\defcitealias{diemer_23_prof2}{II}

\newcommand{\paperone}{Paper \citetalias{diemer_22_prof1}\xspace}
\newcommand{\papertwo}{Paper \citetalias{diemer_23_prof2}\xspace}
\newcommand{\paperonetwo}{Papers \citetalias{diemer_22_prof1} and \citetalias{diemer_23_prof2}\xspace}
\newcommand{\dkft}{\citetalias{diemer_14}\xspace}

\title[Dynamics-based halo density profiles III]{A dynamics-based density profile for dark haloes -- III. Parameter space}

\author[Diemer]{Benedikt Diemer\thanks{Email: \href{mailto:diemer@umd.edu}{diemer@umd.edu}}
\vspace{1mm}
\\
Department of Astronomy, University of Maryland, College Park, MD 20742, USA \\
}

\date{}
\pubyear{2024}

\begin{document}
\label{firstpage}
\pagerange{\pageref{firstpage}--\pageref{lastpage}}
\maketitle


\begin{abstract}
In the previous paper of this series, we proposed a new function to fit halo density profiles out to large radii. This truncated Einasto profile models the inner, orbiting matter as $\rhoorb \propto \exp \left[-2/\alpha\ (r / \rs)^\alpha - 1/\beta\ (r / \rt)^\beta \right]$ and the outer, infalling term as a power-law overdensity. In this paper, we analyse the resulting parameter space of scale radius $\rs$, truncation radius $\rt$, steepening $\alpha$, truncation sharpness $\beta$, infalling normalisation $\delone$, and infalling slope $s$. We show that these parameters are non-degenerate in averaged profiles, and that fits to the total profiles generally recover the underlying properties of the orbiting and infalling terms. We study the connection between profile parameters and halo properties such as mass (or peak height) and accretion rate. We find that the commonly cited dependence of $\alpha$ on peak height is an artefact of fitting Einasto profiles to the actual, truncated profiles. In our fits, $\alpha$ is independent of mass but dependent on accretion rate. When fitting individual halo profiles, the parameters exhibit significant scatter but otherwise follow the same trends. We confirm that the entire profiles are sensitive to the accretion history of haloes, and that the two radial scales $\rs$ and $\rt$ particularly respond to the formation time and recent accretion rate. As a result, $\rt$ is a more accurate measure of the accretion rate than the commonly used radius where the density slope is steepest.
\end{abstract}



\section{Introduction}
\label{sec:intro}

Fitting functions serve a two-fold purpose in astronomy: they provide simplified mathematical descriptions of complex objects or data, and they provide parameter spaces in which we can compare different observations, simulations, and theoretical models. One example is the density structure of dark matter haloes, which takes on enormously complex shapes in detail due to the irregular nature of the cosmic web and due to sub-structure. However, the spherically averaged density profiles of haloes are surprisingly well described by simple formulae such as that of \citet[][hereafter NFW]{navarro_95, navarro_96, navarro_97}, which models profiles as a two-parameter family of mass, $M$, and concentration, $c \equiv R / \rs$. Here, $\rs$ is a scale radius and $R$ denotes some definition of the outer radius (and thus mass) of the halo. A similar parametrisation of a normalization and $\rs$ had been used before, but mostly in the context of describing galaxies \citep{jaffe_83, hernquist_90, dehnen_93, burkert_95}. Critically, NFW showed that concentration is essentially set by the formation epoch of a halo. 

The mass-concentration parameter space has since been the basis of countless investigations, both theoretical \citep[e.g.,][]{bullock_01, wechsler_02} and observational \citep[e.g.,][]{ettori_10}. While the results have thus far confirmed our basic picture of halo structure \citep{umetsu_20_review}, there are also tales of caution. For example, galaxy cluster concentrations measured via weak lensing were persistently higher than expected from simulations, the so-called `over-concentration' problem \citep[e.g.,][]{oguri_12, wiesner_12}. This discrepancy turned out to be due to a correlation in the inferred masses and concentrations \citep{auger_13}, showcasing the dangers of degenerate parameter spaces. 

Moreover, on closer inspection, the $M$-$c$ space turned out not to be quite sufficient to describe halo profiles accurately. The \citet{einasto_65, einasto_69} profile adds a third `steepening' parameter $\alpha$, which describes the rate at which the profile gradually steepens its logarithmic slope. This form describes halos more accurately \citep[e.g.,][]{merritt_06}, even if it is reduced to a two-parameter function by setting the value of $\alpha$ to a constant \citep{wang_20_zoom} or calibrating it as a function of peak height \citep{gao_08, klypin_16}. The chosen parametrisation matters because Einasto concentrations differ from NFW concentrations \citep{dutton_14} and because $\alpha$ and $\rs$ can be degenerate in poorly constrained fits \citep{udrescu_19, eckert_22_einasto}.

Once we push beyond the virial radius, even three parameters are insufficient to describe the profiles in detail because they transition from being dominated by particles orbiting in the halo to being dominated by those falling in for the first time \citep[e.g.,][see, however, \citealt{luciesmith_22_profiles}]{diemand_08}. This infalling term has a much shallower slope, predicted to be roughly $r^{-3/2}$ as long as shells of dark matter do not cross \citep[][see also \citealt{hayashi_08}]{bertschinger_85}. Even accounting for the resulting break, simulated profiles become steeper than the NFW or Einasto models near the transition, which led \citet{baltz_09} to multiply the NFW profile by a cutoff term that depends on a truncation radius, $\rt$, and a power-law exponent.

The steepening has since turned out to be a generic consequence of particle dynamics, namely, a manifestation of the limit of the phase-space distribution of orbiting particles. The position of this `splashback radius' depends mostly on the mass accretion rate (\citealt{diemer_14}, hereafter \dkft; \citealt{adhikari_14}; \citealt{more_15}). \dkft proposed to model the splashback feature by multiplying the Einasto profile by a three-parameter truncation term, $[1+(r/\rt)^\beta]^{-\gamma / \beta}$. The new parameters can be fixed or calibrated as a function of the accretion rate, leaving only the three original Einasto parameters. Nonetheless, the \dkft parameter space suffers from a number of issues. First, $\beta$ and $\gamma$ lack a clear physical meaning because the asymptotic slope of the orbiting profile depends on $\rs$, $\rt$, $\alpha$, and $\gamma$ in a complex fashion. Second, when all parameters are left free, there are significant degeneracies that allow for unphysical parameter values despite reasonable profile shapes \citep{umetsu_17}. This issue becomes particularly relevant when the \dkft profile is used to infer properties of observed profiles, such as the extrapolated shape of the orbiting term \citep{more_16, baxter_17, chang_18, shin_19_rsp, shin_21}.

In this series of papers, we seek to build a better model by understanding the true shape of the orbiting term in and beyond the transition region. In \citet[][hereafter \paperone]{diemer_22_prof1}, we presented a numerical algorithm that distinguishes orbiting and infalling simulation particles according to whether or not they have undergone at least one pericentre. We found that the orbiting term approaches arbitrarily steep slopes at its truncation, and that the profile shapes are affected by accretion rate, mass, and the slope of the linear power spectrum. In \citet[][hereafter \papertwo]{diemer_23_prof2}, we showed that the orbiting profiles are well-described by the form $\rhoorb \propto \exp \left[-2/\alpha\ (r / \rs)^\alpha - 1/\beta\ (r / \rt)^\beta \right]$, i.e., an Einasto profile that is exponentially cut off near the truncation radius (in addition to a power-law infalling profile). One of the main goals was to create a simpler, more meaningful parameter space that directly connects to the physical properties of halos and that has as few degeneracies as possible.

In this paper, we explore this new parameter space by charting the inter-relationships between best-fit parameters, as well as their relationships to mass, accretion rate, and formation time. As in \papertwo, we analyse both individual profiles and averages over halo samples selected by mass and accretion rate. The paper is structured as follows. In Section~\ref{sec:methods}, we briefly describe the underlying simulations, algorithms, halo selection, and fitting functions, but we refer the reader to \paperonetwo for details. In Sections~\ref{sec:res_av} and \ref{sec:res_ind} we present our main results for averaged and individual halos, respectively. We further discuss these results in Section~\ref{sec:discussion} and summarize them in Section~\ref{sec:conclusion}. Supplementary figures are provided online on the author's website at \href{http://www.benediktdiemer.com/data/}{benediktdiemer.com/data}. Throughout the paper, we follow the notation of \paperonetwo. 


\section{Simulations and Methods}
\label{sec:methods}

In this section, we briefly review our simulations and algorithms (Section~\ref{sec:methods:sims}), the selection of halos and the corresponding notation (Section~\ref{sec:methods:halos}), our new fitting function (Section~\ref{sec:methods:func}), and how the best-fit parameters are calculated (Section~\ref{sec:methods:fitting}). We refer the reader to \paperone for details on the simulations, algorithms, and halo definitions, and to \papertwo for details on the new functions and fit quality.

\subsection{N-body Simulations and algorithms}
\label{sec:methods:sims}

We base our analysis on the \erebos suite of $N$-body simulations \citep{diemer_14, diemer_15}, which comprises $14$ simulations of $1024^3$ dark matter particles each. The suite covers different box sizes as well as two \LCDM cosmologies and four self-similar universes. The first \LCDM cosmology is that of the Bolshoi simulation \citep{klypin_11}, consistent with $WMAP7$ \citep{komatsu_11}, namely a flat \LCDM cosmology with $\Omega_{\rm m} = 0.27$, $\Omega_{\rm b} = 0.0469$, $h = 0.7$, $\sigma_8 = 0.82$, and $n_{\rm s} = 0.95$. The second is a \planck-like cosmology \citep[][$\Omega_{\rm m} = 0.32$, $\Omega_{\rm b} = 0.0491$, $h = 0.67$, $\sigma_8 = 0.834$, and $n_{\rm s} = 0.9624$]{planck_14}. The initial power spectra for the respective simulations were generated using \textsc{Camb} \citep{lewis_00}. In contrast, the four self-similar Einstein-de Sitter universes have power-law initial spectra of slopes $-1$, $-1.5$, $-2$, and $-2.5$. These simulations allow us to isolate the impact of the initial power spectrum on the profile parameters \citep[e.g.,][]{knollmann_08, diemer_15, ludlow_17}.

We translate the power spectra into initial conditions using \textsc{2LPTic} \citep{crocce_06}, and we evolve the particle distribution with \textsc{Gadget2} \citep{springel_05_gadget2}. We use the phase--space halo finder \textsc{Rockstar} \citep{behroozi_13_rockstar} to identify haloes and subhaloes, which we connect across cosmic time using the \textsc{Consistent-Trees} code \citep{behroozi_13_trees}. The resulting halo catalogues are described in detail in \citet{diemer_20_catalogs}. 

The novel feature of our analysis is to split the halo density profiles into orbiting and infalling components with the algorithm presented in \paperone. Using the \sparta framework \citep{diemer_17_sparta, diemer_20_catalogs}, we follow the trajectories of each particle in each halo and determine its first pericentre, after which we classify the particle as orbiting rather than infalling \citep[for alternative algorithms, see][]{sugiura_20, garcia_23, salazar_24, enomoto_24}. We have thoroughly tested the convergence and robustness of this algorithm in \paperone, and we refer the reader to that work for details.

\subsection{Halo properties and selection}
\label{sec:methods:halos}

\begin{figure*}
\centering
\includegraphics[trim =  7mm 9mm 1mm 2mm, clip, width=\textwidth]{\figdir/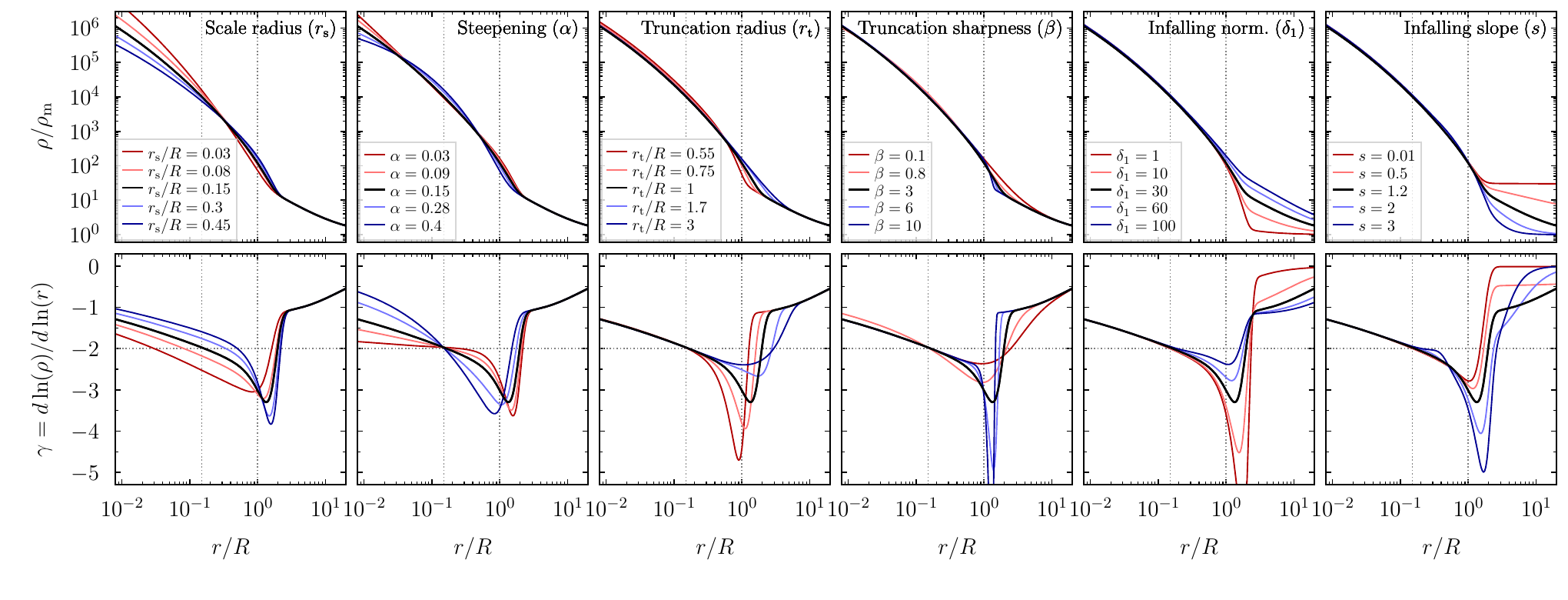}
\caption{Impact of the parameters on our fitting function (Model B plus infalling profile). The reference profile (black line) is kept identical in each panel. We vary one parameter at a time within the ranges of values that we allow in our fits. We omit the fixed parameter $\eta$ as well as $\delmax$, which has no visible influence on the total profile. The normalisation of the orbiting profile, $\rhos$, is adjusted such that all profiles enclose the same mass within $R$. The dotted vertical lines highlight the positions of the reference $\rs$ and $\rt$.}
\label{fig:fitfunc}
\end{figure*}

Wherever possible, we express density profiles in dimensionless units to scale out the absolute size and mass of halos. For this purpose, we normalize the density by the mean matter density of the universe, $\rhom(z)$. The radial dimension can be rescaled by any halo radius $R$ that can be reliably computed for each halo. We use the spherical overdensity radius $\rtom$, which is defined to enclose an average density of $200 \rhom(z)$. The corresponding enclosed mass is $\mtom$, which includes both bound and unbound particles. However, $\mtom$ exhibits well-known dependencies on redshift and cosmology. We thus express mass as peak height, $\nu \equiv \nutom = \deltac / \sigma(\mtom)$, where $\deltac = 1.686$ is the critical density for collapse \citep{gunn_72} and $\sigma$ is the variance of the linear power spectrum. This variance is measured on a scale of the Lagrangian radius of a halo, $R_{\rm L}$, the comoving radius that encloses the halo mass at the mean density of the Universe, $\mtom = (4 \pi/3) \rho_{\rm m}(z=0) R_{\rm L}^3$ (see \paperone). Another variable that has been shown to affect the structure of halos is the logarithmic slope of $\sigma(M)$,
\begin{equation}
\label{eq:neff-1}
\neff(\nu, z) = -2 \left. \frac{d\ln \sigma(R, z)}{d \ln R} \right \vert_{R = R_{\rm L}} - 3 \,,
\end{equation}
which is equivalent to $d \ln(P)  / d \ln(k)$ for scale-free cosmologies with $P \propto k^n$. Thus, $\neff$ represents an effective slope of the power spectrum near the scales relevant for the formation of a halo (its Lagrangian radius). Finally, we define a dimensionless mass accretion rate $\Gamma \equiv \gammadyn \equiv \Delta \ln \mtom / \Delta \ln a$, where $\Delta a$ is one dynamical time defined as the time it takes to cross $2\ \rtom$ (\paperone and \citealt{diemer_20_catalogs}). The accretion over this timescale has been shown to correlate most tightly with halo structure, namely, with the position of the splashback radius \citep{shin_23}.

We consider only host (isolated) halos that are resolved with at least $500$ particles within $\rtom$, and we avoid halos that are experiencing strong interactions with a neighbour by limiting the fraction of unbound mass, $\mtomall / \mtombnd < 1.5$. We measure the radial density profiles between $0.01$ and $10\ \rtom$, but we ignore the inner parts that are affected by the force resolution of the simulation or two-body scattering \citep[Appendix A1 of \paperone;][]{ludlow_19, mansfield_21_resolution, muni_24}. 

As in \papertwo, we fit the density profiles of both individual halos and the average profiles of binned halo samples with certain properties, namely peak height $\nu$, redshift $z$, and optionally accretion rate $\Gamma$. These samples contain halos from all \erebos simulation boxes that have sufficient resolution. In the self-similar universes, there is no physical time so that all redshifts are combined into samples with particular $\nu$ and $\Gamma$. Whenever we plot best-fit parameters as a function of the properties of halo samples, we refer to the median $\nu$ and $\Gamma$ of the halos in a given sample. The effective power spectrum slope corresponding to a halo sample is computed from its median $\nu$ using \eqmn{eq:neff-1}.

\subsection{Fitting function and parameters}
\label{sec:methods:func}

\begin{figure*}
\centering
\includegraphics[trim =  4mm 9mm 0mm 3mm, clip, width=\textwidth]{\figdir/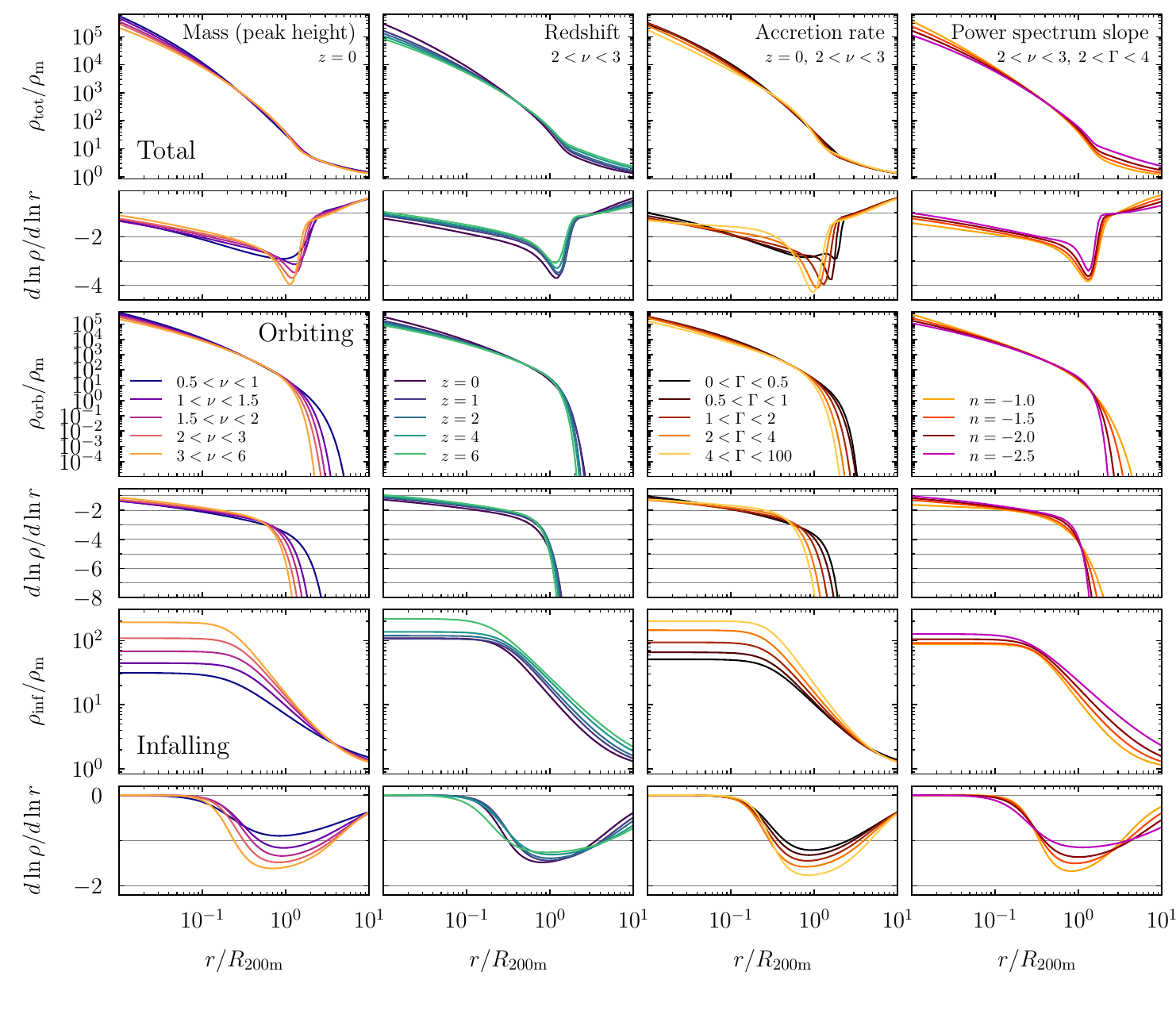}
\caption{Dependence of the best-fit profiles on mass, redshift, accretion rate, and power spectrum slope (from left to right). The large panels show the median total, orbiting, and infalling profiles (from top to bottom), and the smaller bottom panels show the logarithmic slope. Many of the fits shown are also displayed in \papertwo, but here we omit the simulation data to avoid crowding the figure. The profiles are selected to highlight trends in the best-fit parameters. Most notably, changing peak height causes a trend in the scale radius (the concentration-mass relation), and also a trend in the truncation radius. The latter, however, is almost purely driven by the accretion rate, which is responsible for large changes in $\rt$ at fixed $\nu$ (third column). The accretion rate also changes $\alpha$ and the slope of the infalling term, $s$. Redshift barely influences the orbiting term but seems to alter the normalisation of the infalling term, $\delone$ (second column). Physically, this effect is not due to cosmic time but due to changes in the effective slope of the power spectrum, $\neff$. Even at fixed $\nu$ and $\Gamma$, $\neff$ further influences the profiles, as evidenced by the self-similar simulations with different $n$ (right column). Here, higher $n$ leads to power-law like profiles with low steepening $\alpha$, as well as different infalling profiles.
}
\label{fig:fits}
\end{figure*}

Following the insight that the density profile is composed of orbiting and infalling particles with entirely different dynamics, we wrote $\rho(r) = \rho_{\rm orb}(r) + \rho_{\rm inf}(r)$ and suggested new forms for both terms in \papertwo. We built upon the Einasto profile, where the shape of the orbiting term is determined by a normalization $\rhos$ and a radial evolution $S(r)$,
\begin{equation}
\rho_{\rm orb}(r) = \rhos e^{S(r)} \,.
\end{equation}
The Einasto profile sets $S$ such that the logarithmic slope becomes $\gamma \equiv \dnorminl{\ln \rho}{\ln r} = -2 (r / \rs)^\alpha$, i.e., a profile that gradually steepens at a rate $\alpha$ and reaches a slope of $-2$ at the scale radius $\rs$. We extend the Einasto form by a truncation term, 
\begin{equation}
S(r) = -\frac{2}{\alpha} \left[ \rrsa - 1 \right] -\frac{1}{\beta} \left[ \rrtb - \rsrtb \right] \,.
\end{equation}
This profile, which we called Model A, has a logarithmic slope of
\begin{equation}
\gamma(r) = -2 \rrsa - \rrtb \,,
\end{equation}
where the truncation radius $\rt$ sets the location of the edge of the orbiting distribution and $\beta$ determines how sharply the profile truncates. We also presented a slightly more complicated Model B, which gives almost identical profiles but reinstates the meaning of the scale radius as $\gamma(\rs) = -2$,
\begin{equation}
\label{eq:modb}
S = -\frac{2}{\alpha} \left[ \rrsa - 1 \right] -\frac{1}{\beta} \left[ \rrtb - \rsrtb \right] + \frac{1}{\eta} \rsrtb \left[ \rrse - 1 \right] \,,
\end{equation}
where we set the nuisance parameter $\eta = 0.1$ throughout. As shown in \papertwo, Models A and B are indistinguishable across almost all of parameter space, except where $\rs$ approaches $\rt$. In this particular situation, Model B avoids a degeneracy between $\rs$ and other parameters. We thus present Model B results throughout this paper, but we emphasize that the results for Model A are virtually identical (see online figures).

One of the insights of \paperone was that the infalling profile asymptotes to a fixed density at the halo centre. In \papertwo, we found that the infalling term is thus well-described by a power law in overdensity with a normalization $\delone$, a slope $s$, and a maximum overdensity $\delmax$,
\begin{equation}
\label{eq:models:plmk}
\rho(r) = \rhom \left( \frac{\delone}{\sqrt{(\delone / \delmax)^{2} + (r / R)^{2s}}} + 1  \right)  \,.
\end{equation}
In this paper, we mostly focus on fits to the total (orbiting plus infalling) profiles. Fig.~\ref{fig:fitfunc} demonstrates that the interplay between these two components can significantly change the effect a parameter has on the overall profile shape. For example, even if we keep $\beta$ constant, changing $\rt$ produces a cutoff of a different sharpness because the truncation happens on top of a different density of infalling material; the same effect can be observed when changing $\delone$. All parameters change the shape of the truncation (or splashback) feature to some extent. We omit $\delmax$ from Fig.~\ref{fig:fitfunc} since it does not appreciably affect the total profiles. 

Fig.~\ref{fig:fits} shows examples of best-fit profiles to median profiles of halo samples with different masses (peak heights), redshifts, accretion rates, and power spectrum slopes. \paperone showed that the variations in mass and redshift are manifestations of underlying variations with accretion rate and power spectrum slope. The top, middle, and bottom sets of panels show the total, orbiting, and infalling profiles, respectively, highlighting the interplay between these components.

Recently, \citet{salazar_24} introduced an alternative parametrisation for both the orbiting and infalling profiles. Their functions are designed to fit halo samples selected by orbiting mass (rather than $\mtom$, see also \citealt{garcia_23}). Nonetheless, both their form and equation~\ref{eq:modb} fit the orbiting profiles probed by their simulation well, although the two functions predict very different slopes at $r \ll \rs$ (which could not be tested numerically). The \citet{salazar_24} function replaces $\rs$ with a varying slope, leaving a single radial scale akin to $\rt$. Their best-fit parameters are thus not directly comparable to this work. \citet{salazar_24} also point out that the power-law infalling profile (equation~\ref{eq:models:plmk}) is designed to fit the overdensity $\rho_{\rm inf} / \rhom$ within $r \lsim 10\ \rtom$. However, it fails at larger radii, especially when comparing the predicted correlation function, $\xi(r) \propto \rho_{\rm inf} / \rhom - 1$, to simulation data. Matching the correlation function out to large, linear scales demands a much more complex model, as shown by \citet{salazar_24}. We thus caution that equation~\ref{eq:models:plmk} is a purely phenomenological fitting function that should not be extrapolated beyond $10\ \rtom$.

\subsection{Fitting method}
\label{sec:methods:fitting}

Our fitting routine is described in detail in \papertwo, but we briefly summarise it for completeness. We use a least-squares fit to minimise the Cauchy loss function of the logarithmic difference between simulated and fitted profiles,
\begin{equation}
\label{eq:chi2}
\chi_{\rm cauchy}^2 \equiv \sum_i \ln \left(1 + \chi_{\rmi}^2 \right) \quad \mathrm{with} \quad \chi_{\rmi} \equiv \frac{\ln \rho_{\rm i,fit} - \ln \rho_{\rm i,data}}{\ln(1 + \sigma_\rmi / \rho_{\rm i,data})} \,,
\end{equation}
where $i$ indexes the radial bins and $\sigma_\rmi$ is the uncertainty in $\rho_{\rm i,data}$. The Cauchy function reduces the influence of outliers in the fit, which leads to more physically meaningful best-fit parameters. For averaged profiles, $\sigma_\rmi$ is the sum of a bootstrap estimate of the statistical uncertainty and a 5\% systematic error that is added in quadrature in order to avoid low-$\sigma$ bins dominating the fit. For individual halos, the bootstrap estimate is replaced by a Poisson-like statistical uncertainty.

All parameters are translated to log space for fitting. We estimate their uncertainty from the covariance matrix $\vect{C} = (\vect{J}^T \vect{J})^{-1}$, which is in turn computed from the Jacobian matrix returned by the fitting routine. However, the Jacobian refers to changes in the $\chi^2$ measure, whose normalization directly depends on unreliable assumptions about the uncertainties in the profiles (such as the systematic error discussed above). We thus renormalize the covariance matrix and compute the parameter uncertainties from its diagonal,
\begin{equation}
\sigma_{\rm p,j} = \sqrt{\vect{C}_{\rm jj}  \times \chidof} \,,
\end{equation}
where the index $j$ runs over the number of parameters. This renormalization assumes that the `true' $\chidof \approx 1$, i.e., that the fit is reasonably good and that a large or small $\chidof$ must be caused by over- or underestimated errors. Moreover, our expression for $\sigma_{\rm p,j}$ ignores degeneracies between the parameters and should thus be seen as a rough estimate. We show the uncertainties on parameters as gray error bars in the following figures. Routines for fitting the new profile are implemented in the public python package \colossus \citep{diemer_18_colossus}.


\section{Parameters from averaged profiles}
\label{sec:res_av}

In this section, we investigate the best-fit parameters from fits to the mean and median profiles of various halo samples. Our goal is to establish whether the fitting function presented in Section~\ref{sec:methods:func} produces a non-degenerate, physically meaningful parameter space. We investigate the distribution of parameters and potential degeneracies in Section~\ref{sec:res_av:degeneracies}. In Section~\ref{sec:res_av:relations}, we connect the parameters to the physical properties of halos such as mass, accretion rate, redshift, and cosmology. We mostly focus on fits to the total profiles rather than separate fits to the orbiting and infalling terms. While the parameters from the latter are closer to the `true' underlying shapes of the two terms, we only have access to the total profiles in typical applications --- most notably, in observational data. However, in Section~\ref{sec:res_av:total} we show that the total fits generally recover the parameters from the separate fits well. We use model B throughout, but we emphasize that model A gives identical best-fit parameters in the vast majority of cases (\papertwo). 

Fig.~\ref{fig:fits} shows a selection of fits arranged to highlight the dependence of the profiles on halo mass, accretion rate, redshift, and the slope of the power spectrum. This figure will serve as a visual guide while we discuss the respective trends from the perspective of the best-fit parameters.

\subsection{Parameter relations and degeneracies}
\label{sec:res_av:degeneracies}

\begin{figure*}
\centering
\includegraphics[trim =  8mm 8mm 8mm 7mm, clip, width=\textwidth]{\figdir/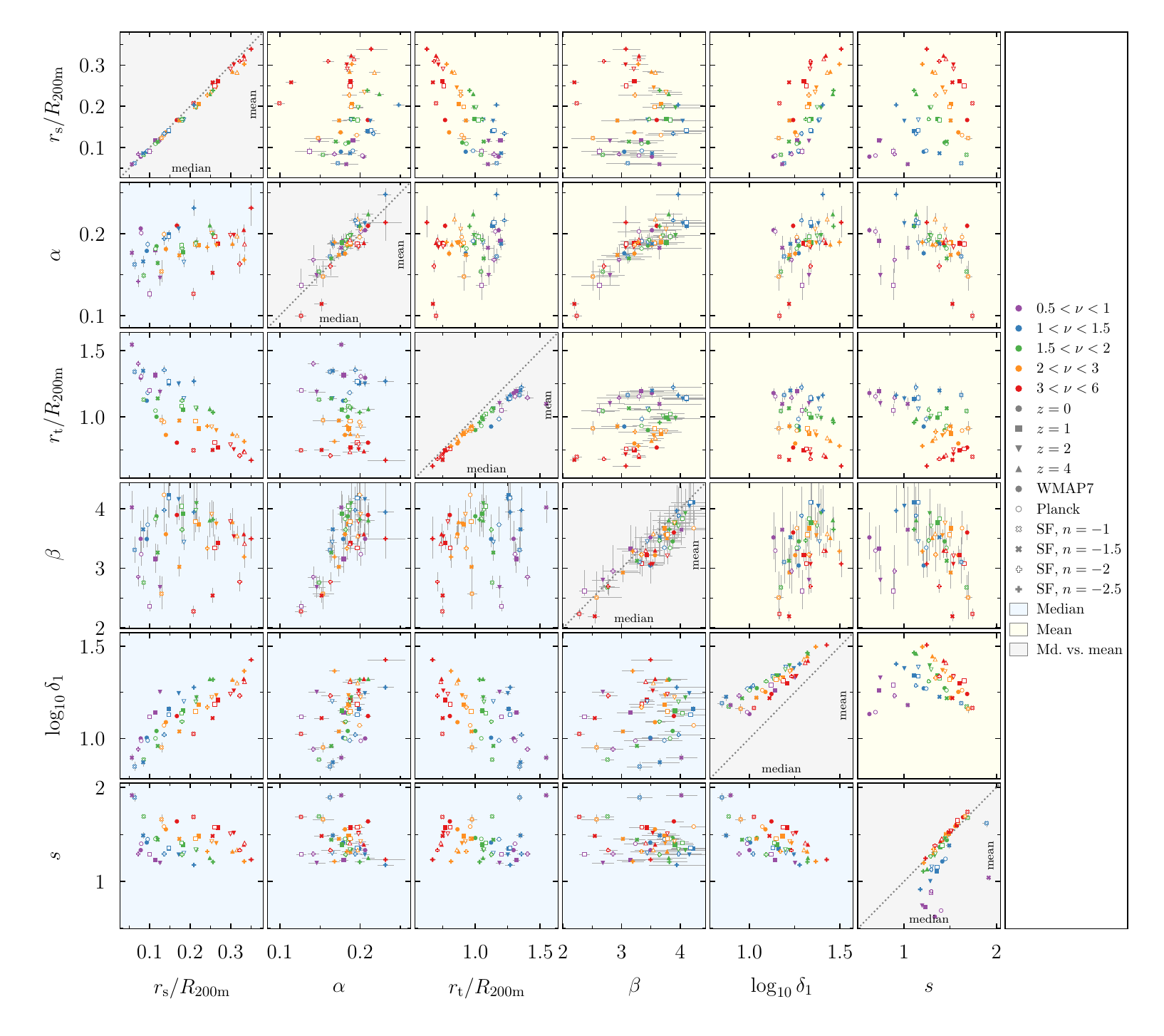}
\caption{Best-fit parameters from fits to the total (orbiting plus infalling) profiles of all mass-selected samples. The bottom-left half (light-blue background) shows the fits to median profiles, and the right-top half (yellow background) mean profiles. The gray diagonal panels compare the median and mean parameters (with median on the $x$-axis). Colours indicate $\nu$ bins, and symbols redshifts as indicated in the legend. Filled symbols correspond to the \wmap cosmology, empty symbols to \planck. The self-similar cosmologies (where all redshifts are combined into a single sample) are shown as filled/empty crosses and pluses. Overall, the parameters from mean and median fits agree well. Wherever parameters fall on tight relations with each other, those sequences are explained by changes in peak height (colour), indicating that none of the profile parameters are strongly degenerate (see Section~\ref{sec:res_av:degeneracies} for details).}
\label{fig:param_nu}
\end{figure*}

Fig.~\ref{fig:param_nu} shows the best-fit parameters from fits to all mass-selected samples, as long as their average profiles are based on at least $80$ profiles (\papertwo). We remove two samples where the fit did not result in converged parameters (low $\nu$ bins in the $n = -1$ cosmology and in \wmap at $z = 4$). Each fit is shown as a set of points with a certain color (peak height) and shape (redshift and cosmology). The uncertainties are shown as gray error bars (Section~\ref{sec:methods:fitting})

The figure is segmented into three areas: the bottom-left triangle (lightly shaded blue) shows the results from fits to median profiles, the top-right triangle (light yellow) those from mean profiles, and the diagonal (gray shaded) panels compare the results from mean and median fits for each parameter. These comparisons are important because the median profiles are less susceptible to outliers and often lead to more well-defined trends in the best-fit parameters, but stacked observations correspond to mean profiles. The mean profiles prefer a systematically slightly lower truncation radius and a higher normalization of the infalling profile, $\delone$, presumably owing to nearby neighbours that can strongly up-scatter the amplitude of the infalling profile. Other than that, the mean and median parameters generally agree well.

Overall, the parameters occupy a large fraction of the allowed space. We immediately notice some correlations between parameters, but those do not necessarily indicate degeneracies in the fit.\footnote{We have omitted $\rhos$ from the figure, which at first sight appears to be highly degenerate with $\rs$. This degeneracy is a consequence of our parametrisation that uses $\rhos / \rhom$ and $\rs / \rtom = 1/\ctom$. From integration, $\rhos / \rhom = 200\ \ctom^3 / (3 f[\ctom])$, where $f(r/\rs)$ is a dimensionless function of order unity (e.g., $f(x) = \ln(1 + x) - x/(1 + x)$ for an NFW profile). Any deviations from this relation indicate that the normalization of the fitted and simulated profiles differ slightly, or that there are slight deviations in $f(r/\rs)$ because our profile is also a function of $r/\rt$.} Instead, many of them are characterized by a sequence in mass (colour), where both parameters change in unison as a function of mass. We note that $\rs$ and $\alpha$ appear more or less uncorrelated, which is a relief given that they can be notoriously degenerate in underconstrained fits. Similarly, $\rt$ and $\beta$ are basically uncorrelated, which indicates that the profiles constrain their values beyond a generic steepening of the slope. Meanwhile, the two radial scales, $\rs$ and $\rt$, are correlated with a clear trend in $\nu$ (colour), which we will explain by the correlation of each with accretion rate. The steepening parameters $\alpha$ and $\beta$ are correlated but positively, which is not what we would expect if they were trading off each other to create profiles of a particular steepness. The normalisation of the infalling profile also falls on sequences in peak height, for example when plotted against the spatial scales $\rs$ and $\rt$. The infalling slope, $s$, however, is more or less uncorrelated with the other parameters. The central infalling overdensity $\delmax$ is omitted from the figure as it has no influence on the total profile. 

We do not observe any strong trends with redshift, which would manifest as squares/triangles occupying different areas than circles ($z = 0$). This non-evolution agrees with our finding that the orbiting profiles are only weakly redshift-dependent (Fig.~\ref{fig:fits}). There is a trend in the normalization of the infalling profile ($\delone$), but it is hard to discern in Fig.~\ref{fig:param_nu}. We also do not notice any obvious trends with cosmology (filled/open symbols). The self-similar cosmologies (crosses and pluses) tend to span a wider range than different \LCDM redshifts, which makes sense given that the redshift evolution in \LCDM cosmologies essentially represents an interpolation of the different power spectrum slopes probed by the self-similar universes (\paperone).

\begin{figure*}
\centering
\includegraphics[trim =  8mm 8mm 8mm 7mm, clip, width=\textwidth]{\figdir/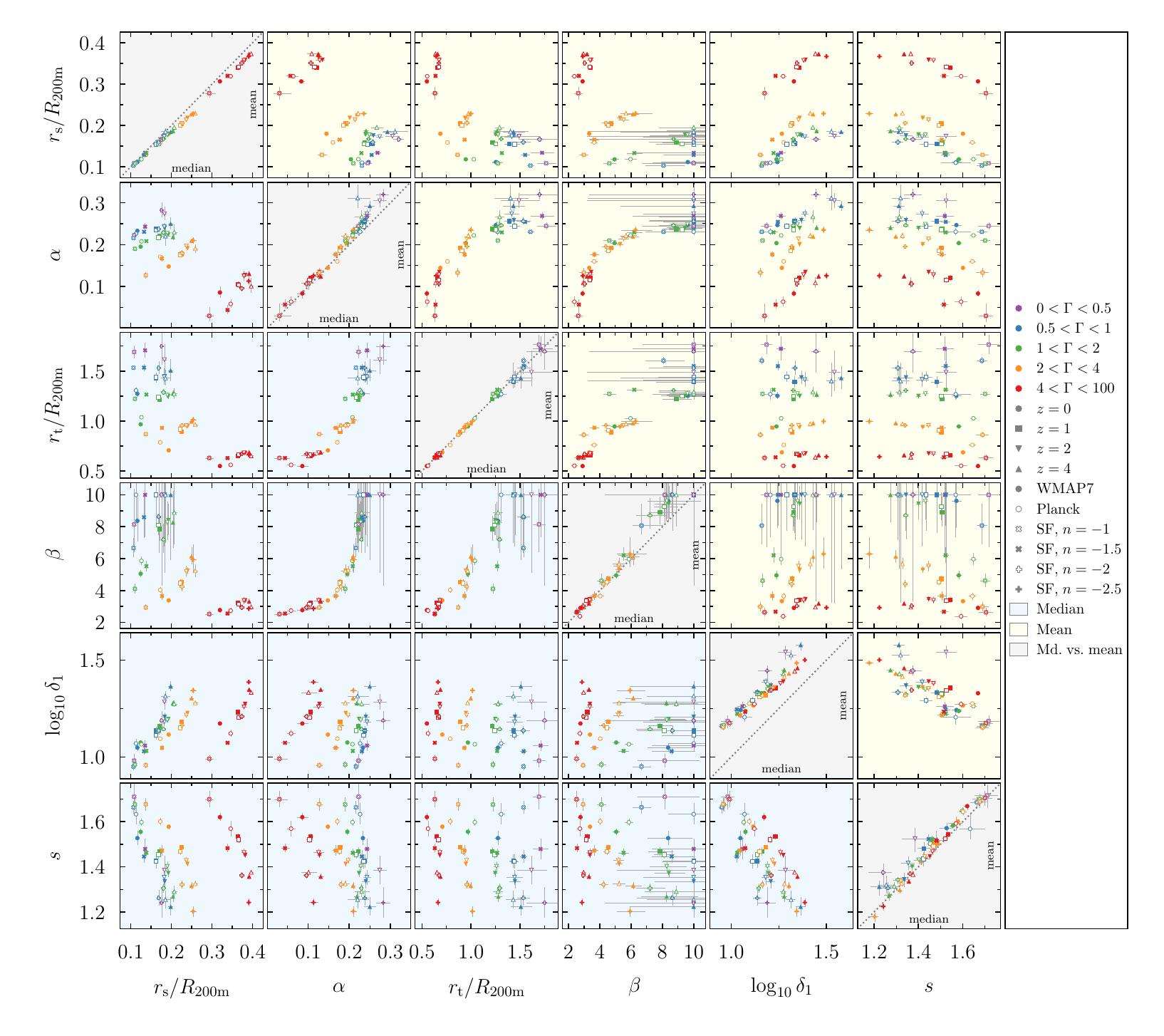}
\caption{Same as Fig.~\ref{fig:param_nu}, but for accretion rate-selected samples with $2 < \nu < 3$. The respective plots for the other $\nu$ bins appear generally similar, although there is more scatter at low and high $\nu$. Many parameters exhibit strong trends with $\Gamma$ (colour), most notably $\alpha$, $\rt$, and $\beta$. As for the $\nu$-selected samples, the mean and median parameters mostly agree, with the exception of the normalisation of the infalling profile, which is sensitive to nearby neighbours.}
\label{fig:param_nugamma}
\end{figure*}

Some of the trends we have observed are readily explained by dependencies on the mass accretion rate, which has a stronger impact on the profiles than mass. Fig.~\ref{fig:param_nugamma} shows the same plot as Fig.~\ref{fig:param_nu}, but for $\Gamma$-selected samples. The point colour now indicates $\Gamma$ rather than $\nu$. We have chosen the $2 < \nu < 3$ bin because the fit parameters are particularly well-determined for high-mass halos, but other $\nu$ bins lead to qualitatively similar conclusions (see online figures). The error bars tend to be larger than for $\nu$-selected samples because the profiles exhibit a greater variety of shapes and are thus harder to fit (\papertwo). 

Overall, the median and mean fits agree excellently for $\Gamma$-selected profiles. We observe essentially no systematic differences in any of the parameters for the orbiting profile, and the infalling slopes $s$ also agree in well-constrained profiles. As in Fig.~\ref{fig:param_nu}, by far the largest differences occur in $\delone$, indicating that a few halos with high infalling profiles increase the mean compared to the median. The relationships between parameters are also clearer in Fig.~\ref{fig:param_nugamma}. Most notably, the parameters that describe the steepening of the profile ($\alpha$, $\beta$, and $\rt$) fall on tight relations that are clearly determined by the accretion rate (rather than by degenerate fits). The anti-correlation of $\rs$ and $\rt$ can easily be explained based on the general shape of formation histories. Concentration is tightly related to the `formation time' when the halo transition from fast to slow accretion \citep{navarro_97, wechsler_02, tasitsiomi_04_clusterprof, zhao_09_acchist, ludlow_13}, whereas $\rt$ captures recent accretion \citep{shin_23}. Specifically, a high accretion rate leads to added matter, contracting orbits, and thus a smaller $\rt$ (or splashback radius, see \citealt{adhikari_14}). These qualitative results are also supported by AI-based decompositions of profiles that are entirely agnostic of any fitting functions \citep{luciesmith_24}. Since halos that formed earlier tend to accrete slowly today, a higher concentration means a lower $\rs$ and a higher $\rt$ (and vice versa). However, the $\rs$-$\rt$ relation maintains roughly the same amount of scatter as for $\nu$-selected samples, highlighting the fundamental two-scale nature of the profiles that cannot be captured by a single radius parameter (see Section~\ref{sec:discussion:mah} for further discussion). 

The truncation tends to be sharper in $\Gamma$-selected samples, leading to much higher values of $\beta$ (although some of the most extreme values are poorly constrained). Moreover, slowly accreting halos (with large $\rt$) exhibit particularly sharp cutoffs (large $\beta$). The infalling profiles are also strongly influenced by $\Gamma$ (Fig.~\ref{fig:fits}), but $\delone$ and $s$ exhibit significant scatter at fixed $\Gamma$ due to their dependence on mass and cosmology.

\subsection{Dependence on halo properties, redshift, and cosmology}
\label{sec:res_av:relations}

\begin{figure}
\centering
\includegraphics[trim =  5mm 10mm 3mm 3mm, clip, scale=0.8]{\figdir/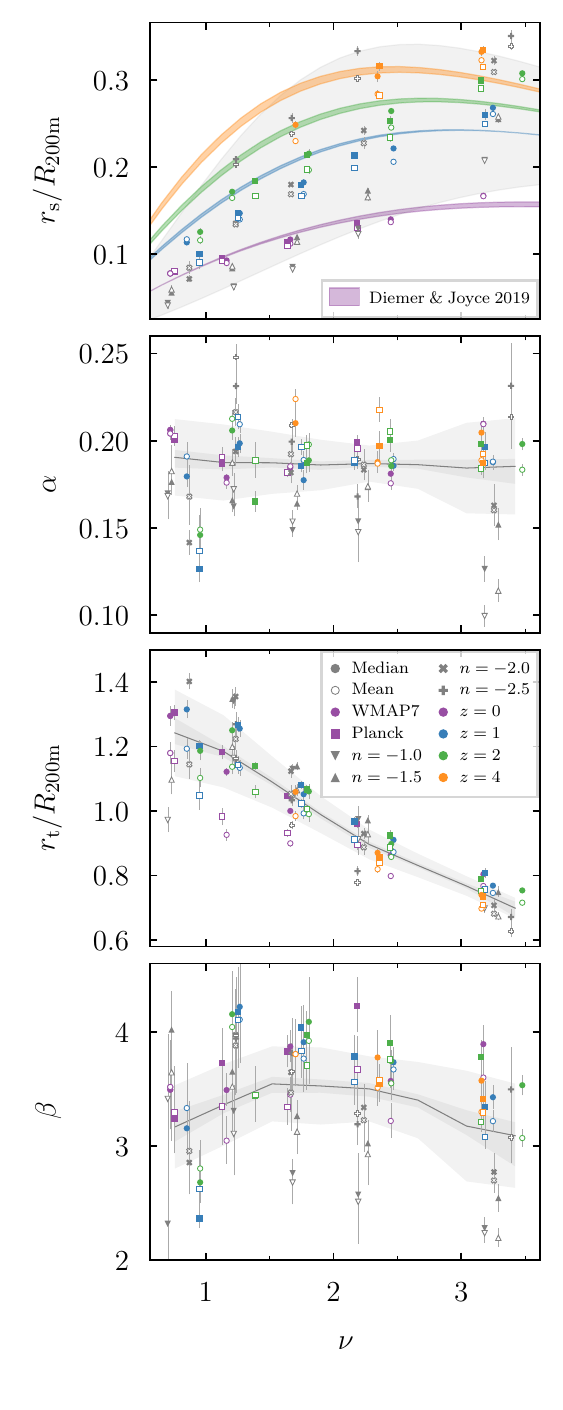}
\caption{Dependencies of the best-fit parameters on peak height. Each point marks a fit to a total profile, with its shape indicating the cosmology, its colour the redshift (gray for self-similar cosmologies), and empty/full symbols indicating fits to mean/median profiles. The gray lines and contours show a kernel-smoothed running median and scatter. The trends can be roughly understood as follows. In the top panel, $\rs / \rtom = 1 / \ctom$ increases with halo mass as expected. The fits roughly match the predictions of \citet{diemer_19_cm}, although their model was calibrated on the NFW concentrations of individual halos. The shaded regions enclose the variations due to cosmology, which are large for the self-similar universes (gray area). The second panel demonstrates that our fits suggest a constant $\alpha$-$\nu$ relation (see further discussion in Section~\ref{sec:res_av:relations:alpha}). The third panel shows that $\rt / \rtom$ (and thus the splashback radius) decreases with $\nu$ because the average mass accretion rate increases. Finally, we find no significant trend of $\beta$ with $\nu$.}
\label{fig:rel_nu}
\end{figure}

\begin{figure}
\centering
\includegraphics[trim =  5mm 10mm 3mm 3mm, clip, scale=0.8]{\figdir/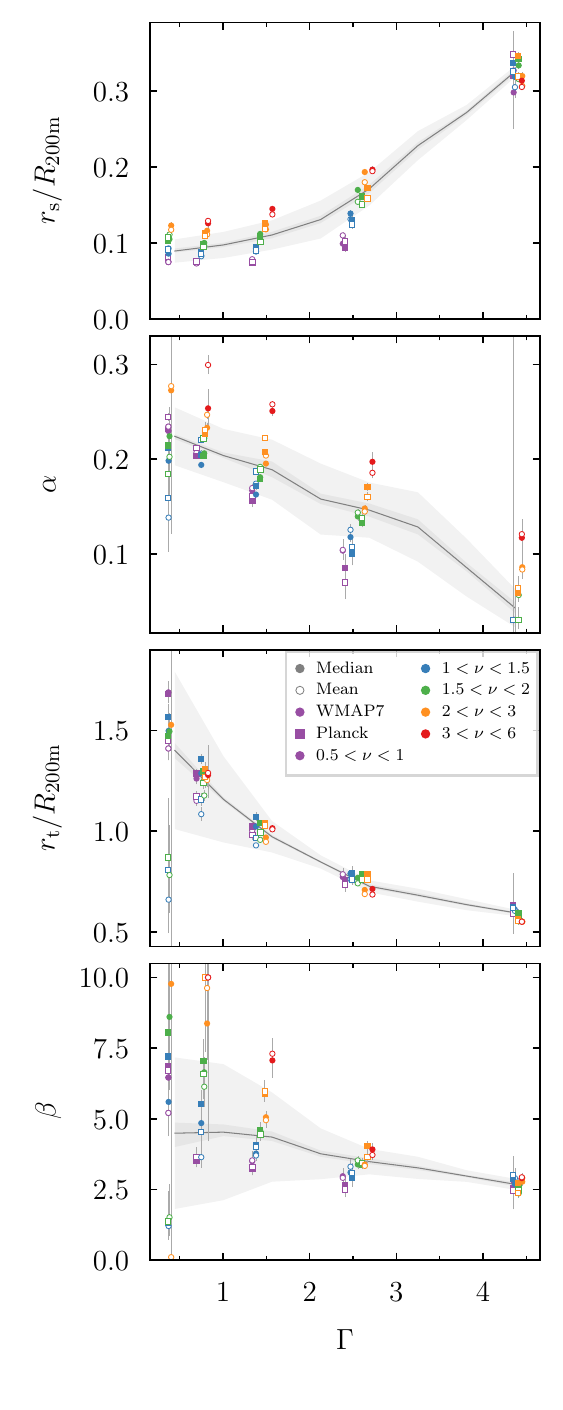}
\caption{Same as Fig.~\ref{fig:rel_nu} but for dependencies on the accretion rate $\Gamma$. The point colours now represent peak height. We omit higher redshifts and the self-similar simulations to avoid crowding, but our conclusions hold regardless. The top panel shows a clear relation between $\Gamma$ and $\rs / \rtom = 1 / \ctom$, which is physically expected since faster-accreting halos are younger and thus have lower concentration. However, the relation is complicated by secondary dependencies on $\nu$ and $z$. Second, $\alpha$ decreases towards high $\Gamma$ to accommodate profiles that increasingly approach power laws (\paperone). The truncation radius (third panel) exhibits a well-defined relation with $\Gamma$. Finally, $\beta$ weakly but noticeably decreases with increasing $\Gamma$. This trend may seem counter-intuitive given that the sharp cutoffs are easier to discern in fast-accreting halos, but it is congruent with the profiles shown in Fig.~\ref{fig:fits}. The total profile is also modulated by the infalling term, which itself depends strongly on $\Gamma$ (Fig.~\ref{fig:rel_gamma_inf}).}
\label{fig:rel_gamma}
\end{figure}

To better understand how the physical conditions of halos translate into their density profiles, we now plot the best-fit parameters of the orbiting profile against halo mass (Fig.~\ref{fig:rel_nu}) and accretion rate (Fig.~\ref{fig:rel_gamma}). The meaning of the symbols and colors has changed compared to Figs.~\ref{fig:param_nu} and \ref{fig:param_nugamma}: filled/open symbols now denote median/mean fits, shape denotes cosmology, and colour denotes redshift in Fig.~\ref{fig:rel_nu} and peak height in Fig.~\ref{fig:rel_gamma}. The latter shows results for $z = 0$, but the conclusions are qualitatively similar for other redshifts (although some of the relations evolve with $z$). 

The gray lines and contours show a smoothed median, its uncertainty, and its scatter as computed by kernel-localized linear regression \citep[KLLR,][]{farahi_18, anbajagane_22_baryonic}. We use a Gaussian filter with a width of 10\% of the $x$-extent of the plotted points. The uncertainty on the points is taken into account when computing the median and scatter. We assign a minimum uncertainty of 1\% of the $y$-extent of the points to avoid numerical issues. The calculation of the weighted scatter can fail, in which case we fall back to the unweighted scatter (which tends to be similar). We now discuss trends in each profile parameter in turn.

\subsubsection{Concentration}
\label{sec:res_av:relations:c}

The top panel of Fig.~\ref{fig:rel_nu} shows the inverse of the concentration-mass relation, since $\rs / \rtom = 1 / \ctom$. We omit the smoothed median from this panel because we expect a well-known redshift dependence of the concentration-peak height relation \citep{prada_12, diemer_15}. Instead, the shaded regions show the predictions of the \citet{diemer_19_cm} model for the $c$-$\nu$ relation at the four redshifts for which points are plotted. The width of the contours represents the differences between the \wmap and \planck cosmologies. The gray contour shows the range predicted for the self-similar cosmologies. We should not expect a perfect match to these predictions because the \citet{diemer_19_cm} model was calibrated to match the concentrations of individual halos rather than those of stacked profiles. Moreover, $c$ was measured by fitting NFW rather than Einasto profiles, which leads to slightly different concentrations \citep[][see further discussion in Section~\ref{sec:discussion:c}]{dutton_14}. Nonetheless, our profiles obey a comparable $c$-$\nu$ relation with a similar redshift trend. As expected from the large gray contour, the self-similar fits scatter across a wide region that mostly encloses the \LCDM results. Fig.~\ref{fig:rel_gamma} reveals that concentration (or $\rs / \rtom$) does correlate significantly with accretion rate in the expected fashion, where young halos in the fast-accretion regime have low concentration \citep[e.g.,][]{wechsler_02, zhao_09_acchist}. Any residual correlation with $\nu$ manifests itself as a colour gradient in the points.

\subsubsection{The steepening parameter $\alpha$}
\label{sec:res_av:relations:alpha}

\begin{figure}
\centering
\includegraphics[trim =  5mm 24mm 3mm 3mm, clip, scale=0.82]{\figdir/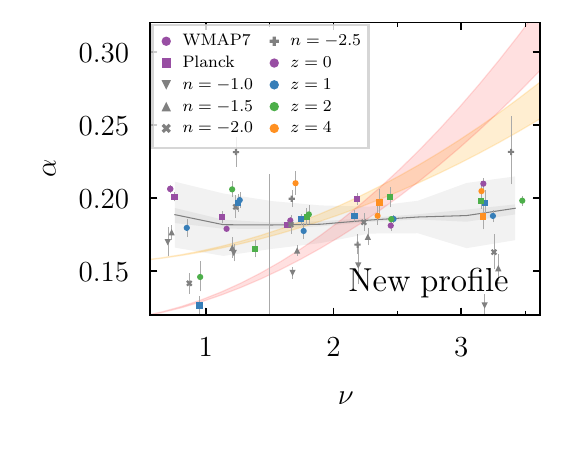}
\includegraphics[trim =  5mm 10mm 3mm 3mm, clip, scale=0.82]{\figdir/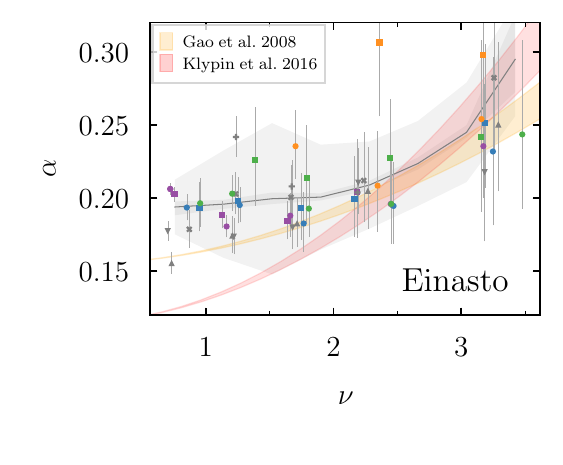}
\caption{The dependence of the profile steepening parameter $\alpha$ on peak height when inferred via our new fitting function (top) or the conventional Einasto profile (bottom). Both forms are combined with a power-law infalling profile. We focus on fits to median profiles. As shown in Fig.~\ref{fig:param_nu}, $\alpha$ shows no evolution with peak height in our new model, which is in tension with the increasing steepening suggested in the literature \citep[][yellow and red shaded areas]{gao_08, klypin_16}. We do, however, recover a similar relation when fitting the Einasto profile. The trend in $\alpha$ arises when fitting profiles that steepen more strongly than the Einasto function naturally predicts. As the profile truncation gets steeper at high $\nu$ (Fig.~\ref{fig:fits}), the best-fit values of $\alpha$ increase to ameliorate a poor fit near the truncation. Since $\rt$ and $\beta$ capture the truncation in our new model, $\alpha$ is determined by the inner profiles ($r \ll \rt$), which prefer a roughly constant $\alpha$.}
\label{fig:rel_alpha}
\end{figure}

In Fig.~\ref{fig:rel_alpha}, we revisit the question of whether the steepening parameter $\alpha$ evolves with peak height. Conventional wisdom has it that $\alpha$ increases with $\nu$, more or less independently of redshift and cosmology \citep{gao_08, dutton_14, klypin_16, udrescu_19, brown_22_einasto}. The fitting functions of \citet{gao_08} and \citet{klypin_16} are shown as yellow and red areas, respectively. They take on a finite width because they were calibrated for $\nutoc$, whose conversion from the $\nutom$ on the $x$-axis depends slightly on redshift and cosmology. Nevertheless, these relations make a clear prediction that $\alpha$ should change from $\approx 0.15$ at low $\nu$ to $\approx 0.25$ at high $\nu$. 

Surprisingly, we find no such trend at all! Instead, the fits scatter around a median value of $\alpha \approx 0.18$ at all peak heights. Fig.~\ref{fig:fits} confirms that there is no indication that the slope steepens more rapidly at small radii for high-$\nu$ than for low-$\nu$ halos. However, when fitting the Einasto profile without the truncation term (bottom panel of Fig.~\ref{fig:rel_alpha}), we qualitatively reproduce the rising $\alpha$ found by other works. This difference demonstrates that the conventional $\nu$-$\alpha$ relation is an artefact of the Einasto profile not quite fitting the steepening slopes near the truncation. While the inner profiles prefer more or less constant values of $\alpha$, higher-$\nu$ halos experience a truncation at smaller radii because of their smaller splashback radii (Fig.~\ref{fig:fits}). Such profiles produce higher best-fit values of $\alpha$ because the fit tries to match the steepening at large radii. The resulting fits may still look acceptable, leading many authors to the conclusion that the Einasto profile produces universally good fits \citep[e.g.,][]{navarro_10, wang_20_zoom, zhou_24}. Moreover, the differences in $\alpha$ will depend on details of the fit such as the relative weighting of bins and the outermost radius.  

Even though $\alpha$ does not evolve with $\nu$ when fitting our new model, it does decrease with accretion rate (second panel of Fig.~\ref{fig:rel_gamma}). This trend arises because of the fairly flat, power-law like profiles of fast-accreting halos (\paperone and Fig.~\ref{fig:fits}). When averaging over different accretion rates to create $\nu$-selected bins, the trend appears to be washed out, leading to the constant $\alpha$ seen in Fig.~\ref{fig:rel_alpha}.

Another result based on Einasto fits is that at fixed $\nu$, $\alpha$ depends on the slope of the power spectrum, $\neff$ \citep{nipoti_15, ludlow_17, brown_20, brown_22_einasto}. Although not obvious from the profiles in the self-similar simulations (Fig.~8 in \paperone), this trend is visible in Fig.~\ref{fig:rel_alpha}, where gray triangles ($n = -1$ and $-1.5$) fall below the median and crosses and pluses ($n = -2$ and $-2.5$) lie above it. The $\neff$-dependence at fixed $\nu$ can be understood in more general terms, as shown by \citet{brown_22_einasto}. The shape of the density profiles, and thus parameters such as $c$ and $\alpha$, depend on the shape of the power spectrum over a wide range of scales. The filter used to compute peak height (e.g., a top-hat) reduces this dependence to a single number, but it does not optimally capture the scales that actually influence $\alpha$, leading to a remaining dependence on the power spectrum that can be summarily parametrised with $\neff$. A similar logic holds for concentration: we can either parametrise it as $c(\nu, n)$ \citep{diemer_15} or as a function of only an altered definition of $\nu$ \citep{brown_22_einasto}.

\subsubsection{The truncation radius and truncation sharpness}
\label{sec:res_av:relations:rt}

We now turn to the location and shape of the truncation. The third panel of Fig.~\ref{fig:rel_nu} shows that our best-fit truncation radii decrease with $\nu$, although with modest variations between $0.6 \lsim \rt \lsim 1.4$. The gray median line in Fig.~\ref{fig:rel_nu} is well approximated as $\rt/\rtom \approx 1.4 - 0.21 \nu$. This relation differs significantly from that of \dkft because $\rt$ takes on a different meaning in their formula for $\nu$-selected profiles (their equation 6). The trend with $\nu$ is almost entirely driven by changes in the average accretion rate, which leads to much stronger variations in $\rt$ (third panel of Fig.~\ref{fig:rel_gamma}). We defer a more detailed analysis of the relation between the truncation radius, accretion rate, and the splashback radius in Section~\ref{sec:discussion:rt}.

Finally, $\beta$ does not vary systematically with $\nu$, which is congruent with the profiles shown in Fig.~\ref{fig:fits}. The changes in profile shape are clearly driven by $\rt$, which does significantly decrease with $\nu$. We do observe a slight decrease of $\beta$ with $\Gamma$, but it is not clear how significant this trend is given the large scatter at low $\Gamma$. Moreover, Fig.~\ref{fig:param_nugamma} shows that $\beta$ and $\alpha$ are highly correlated. These findings raise the question of whether $\beta$ really describes a physical reality or whether it is a `nuisance' parameter. We have experimented with fixing $\beta \sim 4$ and find acceptable fits to most total profiles. However, many of the stacked orbiting terms are fit significantly less well, especially for the mean profiles. While the scatter in the $\Gamma$-$\rt$ relation is reduced, the median does not change significantly. Similarly, the parameters of the inner profile ($\rs$ and $\alpha$) do not qualitatively change. We thus conclude that $\beta$ may not be directly connected to an obvious physical halo property, it does seem to capture a truncation shape that does genuinely vary between halo samples.

\subsubsection{The infalling profile}
\label{sec:res_av:relations:inf}

\begin{figure}
\centering
\includegraphics[trim =  5mm 10mm 3mm 3mm, clip, scale=0.8]{\figdir/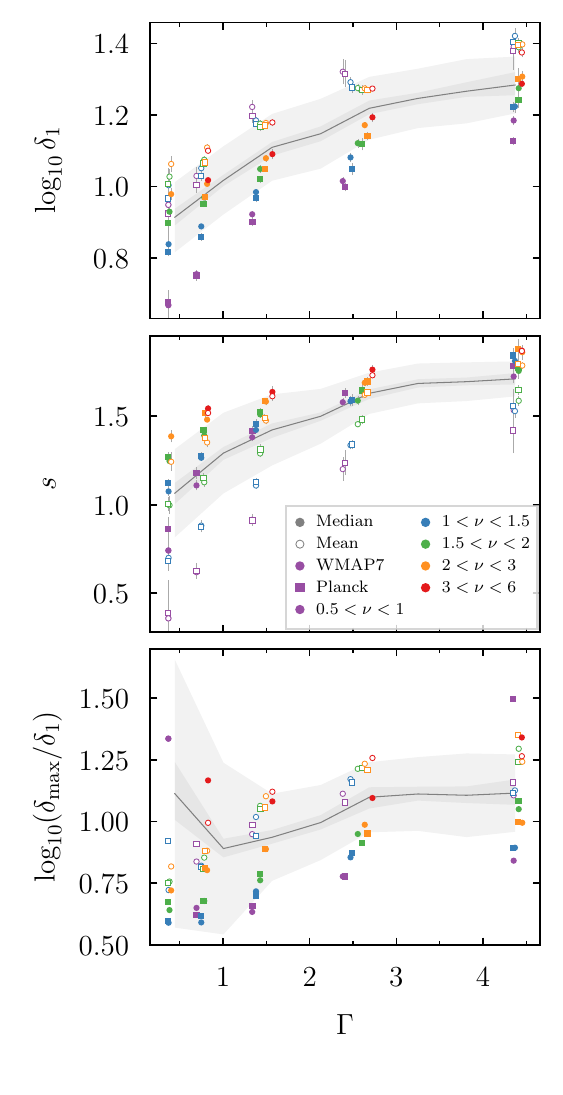}
\caption{Same as Fig.~\ref{fig:rel_gamma} but for the infalling profile. We show parameters based on fits to only the separate infalling profiles because the relations are somewhat less clear when fitting the total profiles (see online figures). The infalling profiles of fast-accreting halos exhibit a higher normalization (top panel), a steeper slope (middle panel), and approach a higher asymptotic density at the halo centre (bottom panel, which shows the more or less constant ratio between $\delmax$ and $\delone$). All parameters show a clear dependence on whether we fit the median or mean profiles (filled or empty symbols), highlighting the asymmetric distribution of infall densities around the mean.}
\label{fig:rel_gamma_inf}
\end{figure}

\begin{figure*}
\centering
\includegraphics[trim =  13mm 24mm 7mm 7mm, clip, width=\textwidth]{\figdir/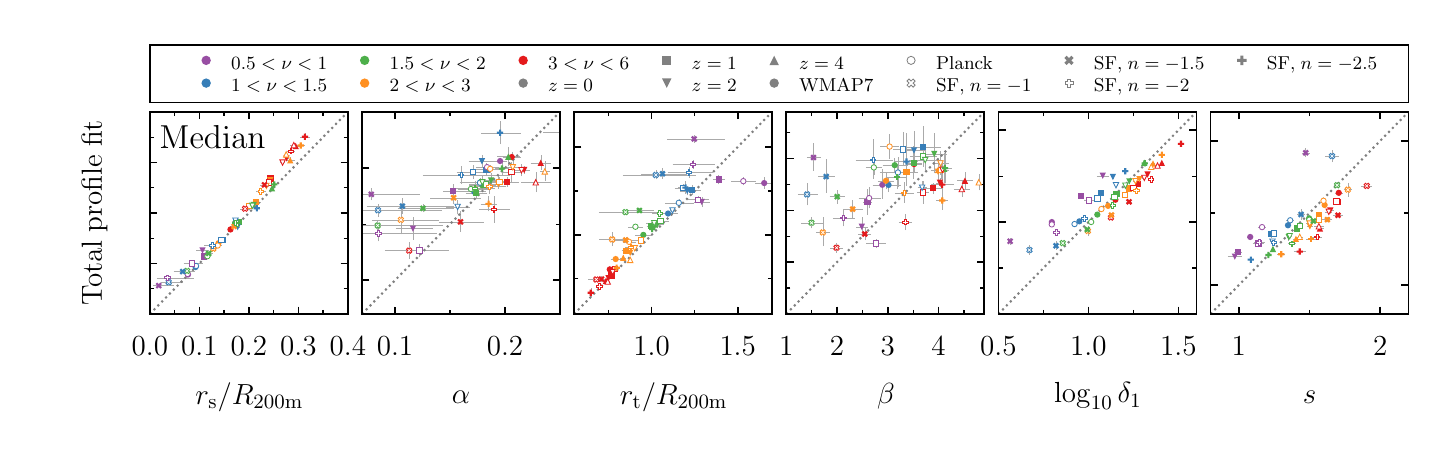}
\includegraphics[trim =  13mm 9mm 7mm 18mm, clip, width=\textwidth]{\figdir/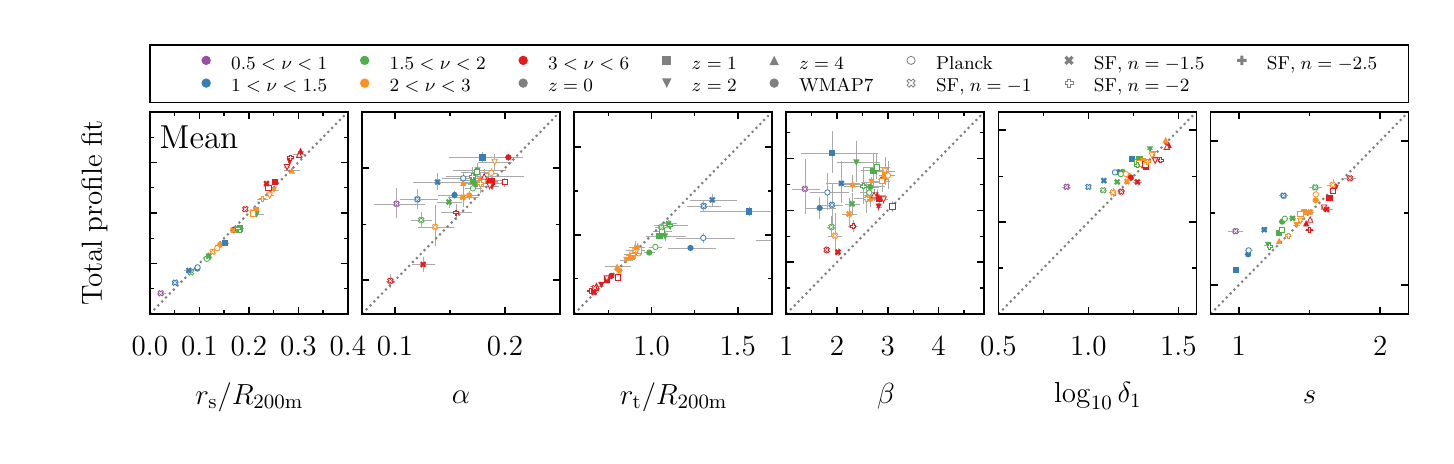}
\caption{Recovery of best-fit parameters from separate fits to the orbiting and infalling terms ($x$-axis) by fits to the total profile ($y$-axis). The two rows show results from fits to the median (top) and mean (bottom) profiles of $\nu$-selected samples. The symbols have the same meaning as in Fig.~\ref{fig:param_nu}. The good agreement for $\rs$ is expected because the inner region of halos is strongly dominated by the orbiting term, but the truncation radius $\rt$ can be strikingly well recovered from total fits. The exception are low-mass haloes (purple/blue), which do not exhibit a strong steepening feature and are less relevant observationally. The agreement is less clear for the steepening parameters $\alpha$ and $\beta$, especially in the mean profiles. However, these parameters also carry somewhat less physical meaning and can be fixed. The parameters of the infalling profile are recovered reliably, especially in high-mass haloes. See Section~\ref{sec:res_av:total} for details.}
\label{fig:totsep}
\end{figure*}

Fig.~\ref{fig:fits} demonstrates that the infalling shapes vary systematically with $\Gamma$ and $\neff$, whereas much of the dependence on $\nu$ is due to the $\nu$-$\Gamma$ relation and nearby neighbours (\paperone). Fig.~\ref{fig:rel_gamma_inf} shows the dependence of the infalling profile parameters on $\Gamma$. Here, we consider the results of a separate fit to the infalling profile only because the trends are somewhat less clear in the total fits, and because $\delmax$ is kept fixed in the total fit.

Perhaps unsurprisingly, the normalization of the infalling profile is higher in fast-accreting halos. If material falls in with some characteristic velocity, a higher flow rate implies more infalling material \citep[e.g.,][]{deboni_16}. Unlike the orbiting profile, the infalling profile is subject to a strong redshift trend: the median normalization $\delone$ increases by about 60\% between $z= 0$ and $z = 4$ (see also Fig.~6 in \paperone), but this increase can be understood as an effect of $\neff$ rather than cosmic time (\paperone).

The slope $s$ becomes steeper with $\Gamma$, approaching $s \approx 1.5$ at high accretion rates, which is the value predicted for non-crossing, collapsing shells \citep[][see \dkft for a similar result]{bertschinger_85}. This picture should break down around the splashback radius, which occurs at larger radii for low-$\Gamma$ halos. Once the mass interior to infalling particles decreases as they move inwards, their density profile flattens (see discussion in \paperone). It is possible that this mechanism causes the decrease in $s$ at low $\Gamma$, but we leave it to future work to understand the reasons in detail. 

The asymptotic maximum density at the halo centre increases with $\Gamma$, but this trend is perhaps easier understood by considering the ratio $\delmax / \delone$ as shown in the bottom panel of Fig.~\ref{fig:rel_gamma_inf}. The relatively constant ratio suggests that the infalling profile reaches a maximum that is a fixed multiple, about ten times, its normalization at $\rtom$. This limit could be related to the typical fractions of radial and tangential particle orbits, but we defer a more detailed investigation to future work. Some outlier points correspond to fits where the central density is unconstrained and takes on its maximum value of $10^4\ \rhom$.

\subsection{Recovering parameters from the total profile}
\label{sec:res_av:total}

Our insights into the separate orbiting and infalling profiles are only meaningful if they can be recovered by fits to the total profiles, which are available observationally and in unsplit simulation data. In this section, we investigate how accurately the two sets of parameters track each other, or, in other words, how much knowledge about the profiles is lost in the superposition of orbiting and infalling matter. Fig.~\ref{fig:totsep} shows the relationship between the best-fit parameters from the separate orbiting/infalling fits (on the $x$-axis) and from the total fits (on the $y$-axis). We consider $\nu$-selected samples, but the conclusions hold for $\Gamma$-selected samples. We omit $\rhos$ because the correspondence is very good as expected, as well as $\delmax$, which is not varied in the total fits. We consider both median profiles (top row) because they show the clearest trends in fit parameters and mean profiles (bottom row) because they correspond most closely to stacked observations. We do not find strong differences between median and mean profiles in Fig.~\ref{fig:totsep}.

Most importantly, the radial scales $\rs$ and $\rt$ are both recovered reliably. While the tight correlation between total and separate fits might be expected at small, orbiting-dominated radii such as $\rs$, it is a positive surprise for $\rt$, which measures the position of the transition region that is by construction most affected by the superposition of orbiting and infalling matter. At low mass (blue/purple points in Fig.~\ref{fig:totsep}), the correlation in $\rt$ degrades, mostly because of erratic fits to separate profiles that contain significant contributions from neighbours (\paperone). This issue is of little importance because observations will, for the foreseeable future, focus on group and cluster haloes.

Recovering the steepening parameters $\alpha$ and $\beta$ is much harder, especially at low mass. Here, the infalling term conceals the sharpness of the truncation that is captured by $\beta$. Both parameters are biased high in the total fits and are subject to significant scatter. However, those parameters convey less physical meaning than $\rs$ and $\rt$, and they can be set to fixed values or limited by an informative prior (\papertwo). The parameters of the infalling profiles, $\delone$ and $s$, are broadly recovered, especially at high $\nu$. 

In summary, the total profiles contain crucial information about the orbiting and infalling terms and the transition region. We note that the fitting procedure matters for these results. For example, by down-weighting outlier points, the Cauchy loss function (Section~\ref{sec:methods:fitting}) leads to much more accurate estimates of the underlying profile parameters than a standard $\chi^2$ metric.
  

\section{Parameters from individual haloes}
\label{sec:res_ind}

\begin{figure*}
\centering
\includegraphics[trim =  1mm 9mm 2mm 2mm, clip, scale=0.72]{\figdir/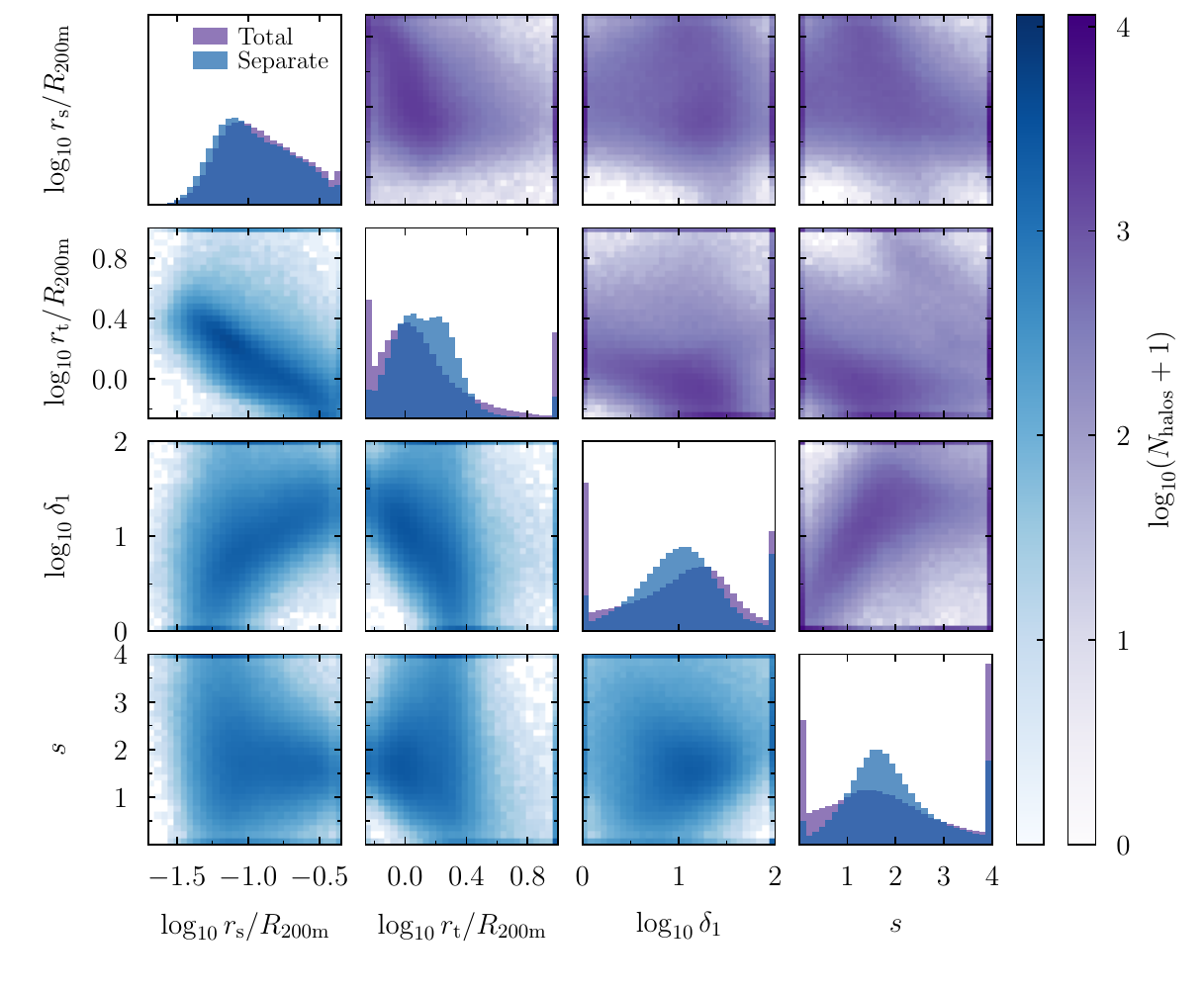}
\caption{Parameter distributions (diagonal panels) and inter-relations (off-diagonal panels) of the best-fit parameters for individual profiles. Since $\alpha$ and $\beta$ are fixed, we show only the scale and truncation radii as well as the normalization and slope of the infalling profile. As before, we omit the normalization $\rhos$ due to its trivial relationship with $\rs$. The bottom-left triangle (blue) shows the results from separate fits to orbiting and infalling profiles, which capture the `true' shape of the underlying profiles (although there are caveats as discussed in the text). In the top-right triangle (purple), we compare to the less constrained, but more commonly available fits to the total profiles. The parameters fill most of the available space, showing that they are not strongly degenerate. The correlation between $\rs$ and $\rt$ is easily discernible in the separate fits (as in Figs.~\ref{fig:param_nu} and \ref{fig:param_nugamma}) but somewhat washed out in the total fits. A significant number of total profiles exhibit no clear truncation, meaning that $\rt$ tends to its lowest or highest allowed values (purple histogram in second diagonal panel). Similarly, the parameters of the outer profile are less reliable in total fits because $\delone$ and $s$ develop a degeneracy that is not present in the separate fits.}
\label{fig:ind_par}
\end{figure*}

The results presented so far have benefited from averaging the density profiles of numerous (at least $80$) halos, which removes stochastic noise that arises from substructure, anisotropies such as filaments, and the finite number of particles in simulations. In this section, we ask whether the trends we found for averaged profiles can be extracted from the noisy distributions of individual fit parameters. On some level, the averaged profiles must reflect the individual profiles they are based on, but due to asymmetric scatter and the complex translation from profiles to best-fit parameters, the properties of mean/median profiles are not always the same as the mean/median properties of individual profiles.

As a result, fits (within the virial radius) are typically restricted to two free parameters, e.g., normalization and scale radius. Even adding a third parameter, such as Einasto's $\alpha$, leads to severe degeneracies \citep{udrescu_19}. Thus, it is hopeless to fit our new model to individual profiles with all seven free parameters (five for the inner and two for the outer profile). Instead, we fix $\alpha = 0.18$ and $\beta = 3$, which leaves $\rhos$, $\rs$, and $\rt$ as the free parameters of the orbiting profile (\papertwo). The chosen values for $\alpha$ and $\beta$ provide a good fit to the vast majority of profiles. The preference for shallow profiles (low $\alpha$) in fast-accreting halos (Fig.~\ref{fig:rel_gamma}) is hard to discern in individual haloes.

We consider both separate fits to the orbiting and infalling profiles as well as total fits, where we fix $\delmax$ to its value from the orbiting-only fit but let $\rhos$, $\rs$, $\rt$, $\delone$, and $s$ vary. The separate fits tell us about the `true,' underlying parameters of the orbiting and infalling components, whereas the total fits tell us what information persists in the sum of these terms. In most figures, we show the entire sample of about \num{378000} fitted halo profiles (\papertwo). The trends in sub-samples (such as \LCDM at $z = 0$) tend to be slightly sharper but not fundamentally different (see online figures).

\subsection{Parameter relations and degeneracies}
\label{sec:res_ind:degeneracies}

Fig.~\ref{fig:ind_par} shows the relationships between fit parameters from individual halos, analogous to Fig.~\ref{fig:param_nu} but omitting the fixed $\alpha$ and $\beta$ parameters. Whereas Fig.~\ref{fig:param_nu} showed results from the median and mean profiles in its lower and upper triangles, we now show parameters from the separate (blue) and total fits (purple). The diagonal panels compare the distributions of each parameter. Both $\rs$ and $\rt$ are shown in log space, and the range for $\rs$ is slightly cut off at its lower end because only a negligible fraction of fits return $\log_{10} \rs/\rtom \lsim -1.7$.

The parameters from the separate fits (blue) give us an impression of the true, underlying distribution of profile properties. The parameters fill a large fraction of the available parameter space, meaning that they are not degenerate. The $\rs$-$\rt$ anti-correlation is still clearly observable in individual haloes, and the significant scatter reminds us that both scales contain separate information about the formation time and accretion rate of the haloes (see further discussion in Section~\ref{sec:discussion:mah}). The other panels show weak correlations similar to those in the averaged profiles (Fig.~\ref{fig:param_nu}): $\delone$ correlates positively with $\rs$ and negatively with $\rt$ (indicating a connection with $\Gamma$ that we discuss below), and $s$ is more or less uncorrelated with the other parameters. 

\subsection{Recovering parameters from the total profile}
\label{sec:res_ind:total}

\begin{figure}
\centering
\includegraphics[trim =  5mm 3mm 2mm 2mm, clip, scale=0.6]{\figdir/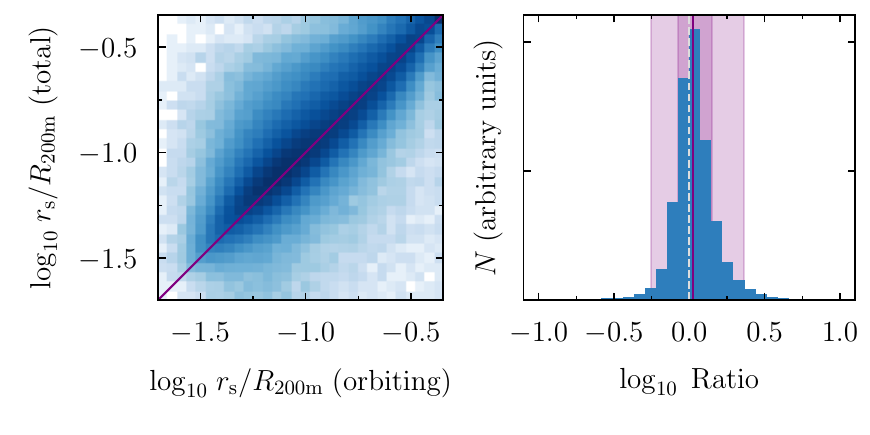}
\includegraphics[trim =  5mm 6mm 2mm 0mm, clip, scale=0.6]{\figdir/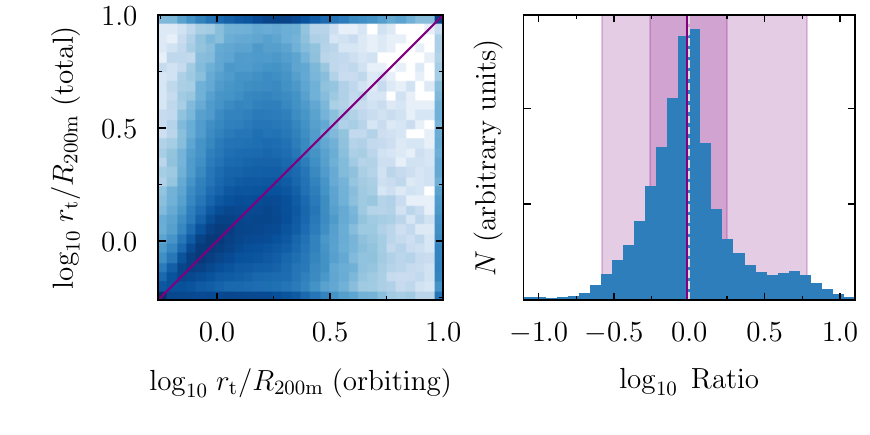}
\caption{Recovery of $\rs$ and $\rt$ from fits to the total profile of individual halos, compared to fits to only the orbiting profile. The left panels show the density on a logarithmic scale similar to Fig.~\ref{fig:ind_par}, whereas the right panels show histograms of the logarithmic ratio. Both $\rs$ and $\rt$ are recovered without bias but with significant 68\% scatter of $0.15$ dex in $\rs$ and $0.25$ dex in $\rt$.}
\label{fig:ind_rs_rt}
\end{figure}

The distributions of the total fit results (purple panels in Fig.~\ref{fig:ind_par}) clearly differ from the separate fits. For example, the $\rs$-$\rt$ relation is weaker, the relations between the orbiting parameters and $\delone$ have changed, and $\delone$ and $s$ are now visibly degenerate. These changes also manifest in the one-dimensional parameter distributions, which exhibit significant fractions of parameters in the lowest and highest allowed bins ---  a sign of poorly converged fits. These challenges are not surprising, given that we fit the profiles with five free parameters. While the infalling profile at $r \gg \rtom$ adds information, this part is also subject to large scatter (\paperone). For example, we observe some cases of density increasing with radius in the infalling part, which violates our requirement that $s \geq 0.01$ (\papertwo). We conclude that it is difficult to extract reliable parameters for the infalling profile from individual halos.

The most interesting parameters, however, are $\rs$ and $\rt$, which tell us about the structure of the orbiting term and the accretion history of the halo (Section~\ref{sec:res_ind:relations}). In Fig.~\ref{fig:ind_rs_rt}, we directly investigate how well the total fits ($y$-axis) recover the values from the individual fits ($x$-axis). The right panels show histograms of the logarithm of the ratio between the two values. In both cases, the total estimates are essentially unbiased, but with 68\% scatter of about $0.15$ dex in $\rs$ and $0.25$ dex in $\rt$. From visual inspection of the fits, we find that large differences are typically due to unusually shaped profiles where our function (or any function) is a poor fit. In such cases, the fit sometimes determines a sensible $\rs$ at the expense of an $\rt$ that poorly represents the truncation, and vice versa.

In summary, individual profiles do contain information about both $\rs$ and $\rt$, and thus about the formation time and recent accretion rate of halos. However, noise in the profiles means that the uncertainty on any individual parameter constraint is significant. These conclusions do not change if we consider sub-samples, such as \wmap at $z = 0$.

\subsection{Dependence on halo properties}
\label{sec:res_ind:relations}

\begin{figure}
\centering
\includegraphics[trim =  8mm 10mm 4mm 4mm, clip, scale=0.71]{\figdir/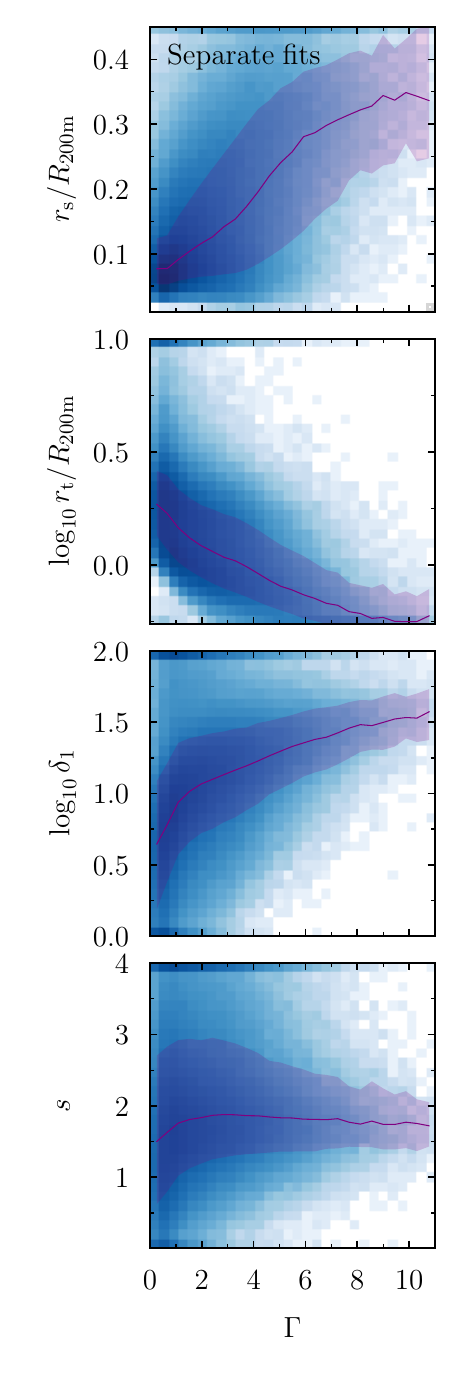}
\includegraphics[trim =  23mm 10mm 4mm 4mm, clip, scale=0.71]{\figdir/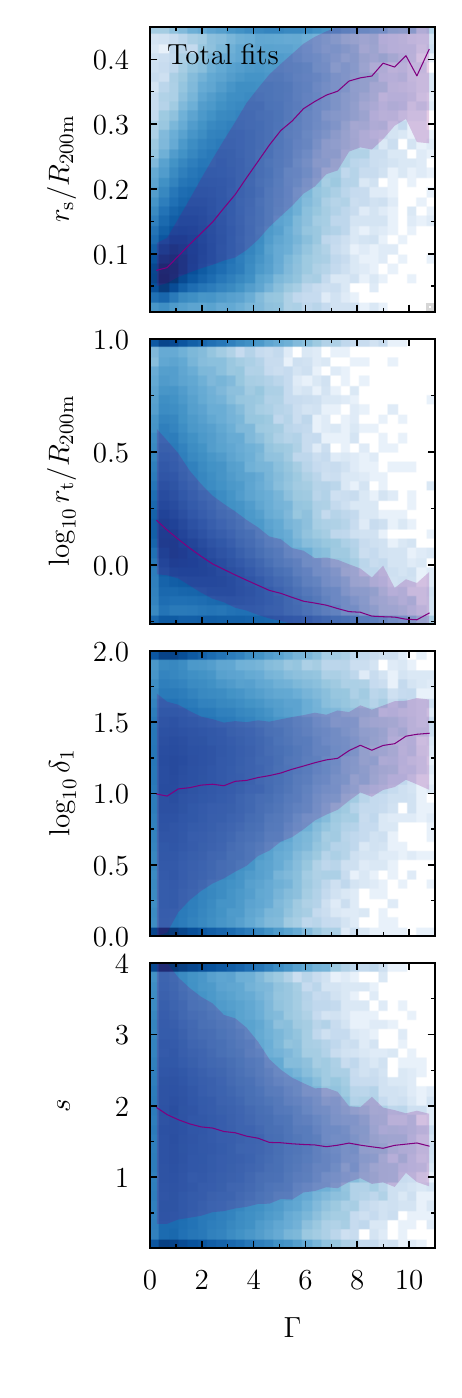}
\caption{Relations between fit parameters and mass accretion rate for individual halos. The density of halos is shown as a logarithmic color map similar to Fig.~\ref{fig:ind_par}. The purple lines and contours show the median and 68\% interval. As in Fig.~\ref{fig:ind_par}, we show results from fits to the separate orbiting and infalling profiles (left) and to the total profiles (right). As expected, the scatter is larger in the latter case for most parameters, especially for the normalization and slope of the outer profile. However, the median relations are remarkably stable to the fitting uncertainty introduced by fitting the total profiles. Most notably, the $\rs$-$\Gamma$ relation is slightly tighter when fitting total profiles (see Section~\ref{sec:res_ind:total}). The median $\rt$-$\Gamma$ relation is virtually identical to that inferred from orbiting-only fits, demonstrating that it is, in principle, possible to retrieve accretion rates from individual profiles, albeit with large scatter.}
\label{fig:ind_rel}
\end{figure}

In Section~\ref{sec:res_av:relations}, we found that the accretion rate is a powerful predictor of the best-fit parameters. We now ask to what extent these relationships can be discerned in noisy individual profiles. Fig.~\ref{fig:ind_rel} recreates Figs.~\ref{fig:rel_gamma} and \ref{fig:rel_gamma_inf}, but now showing the distribution of individual fits (on roughly the same colour scale as Fig.~\ref{fig:ind_par}) as well as its median and scatter (purple lines and shaded regions). We again consider both separate fits (left column) and the more realistic case of total fits (right). The trends are similar to those of averaged profile parameters in Figs.~\ref{fig:rel_gamma} and \ref{fig:rel_gamma_inf}: $\rs$ increases with $\Gamma$ because fast-accreting halos are young and have low concentration; $\rt$ decreases with $\Gamma$; the infalling profile is more prominent in fast-accreting halos (higher $\delone$); and the slope $s$ is more or less uncorrelated with $\Gamma$, in contrast to the averaged profiles. The latter observation indicates that the slight $\Gamma$-$s$ correlation in Fig.~\ref{fig:rel_gamma_inf} might be spurious in the sense that it arises from averaging over infalling profiles with different slopes.

Interestingly, the 68\% scatter in the $\rt$-$\Gamma$ relation is increased only mildly in the total fits (right column), indicating that a meaningful constraint on $\Gamma$ could, in principle, be derived from individual halos. However, the rate of catastrophic outliers is much larger than for the separate fits. The $\rs$-$\Gamma$ relation is tighter in the total fits, which is again explained by the fact that the separate fits are sometimes drawn to the correct $\rt$, at the expense of $\rs$ (Section~\ref{sec:res_ind:total}). We have also checked whether the ratio $\rt / \rs$ (which one could understand as the `truncation concentration') contains separate information from $\rs / \rtom$ and $\rt / \rtom$. While this ratio can reach large values ($\gsim 20$--$30$) at low $\Gamma$, we find that it correlates less with $\Gamma$ than $\rt / \rtom$, indicating that the conventionally defined concentration and $\rt$ are cleaner ways to infer the overall and recent accretion histories \citep[see also][]{shin_23}.

Splitting the overall halo sample by redshift and/or cosmology makes the correlations in Fig.~\ref{fig:ind_rel} slightly tighter but does not qualitatively alter them. The consistency between samples once again demonstrates that accretion history is the most important determinant of the parameters, with only minor influences from peak height and cosmology. We have checked that the relationships of the parameters with $\nu$ are similar to those of averaged profiles shown in Fig.~\ref{fig:rel_nu}, and that they mostly arise from the positive correlation between $\nu$ and $\Gamma$ (see online figures). 


\section{Discussion}
\label{sec:discussion}

We have analysed the best-fit parameters derived from fits to averaged and individual halo profiles, and established how they are connected to halo mass, accretion rate, redshift, and cosmology. In this section, we further discuss the connection to the overall accretion history (Section~\ref{sec:discussion:mah}), how concentration differs between models (Section~\ref{sec:discussion:c}), and how the truncation radius is connected to other definitions of the halo boundary (Section~\ref{sec:discussion:rt}).

\subsection{The connection to a halo's accretion history}
\label{sec:discussion:mah}

Throughout this series of papers, we have alluded to the tight connection between a halo's formation history and its density profile. In the `inside-out' picture of halo formation, shells of dark matter accrete onto a halo seed, and the particle orbits remain at roughly the same (average) radii \citep[e.g.,][]{bertschinger_85}. Though simplified, this picture holds in principle, and halo profiles can be understood as a reflection of the critical density of the Universe at the times when shells were accreted \citep{ludlow_13, correa_15_b, lopezcano_22}. The time when a halo transitions from fast to slow accretion (which we shall identify with $\ahalf$, the scale factor when the halo had accreted half its current $\mtom$) imprints itself onto the profile as the scale radius \citep{wechsler_02, zhao_09_acchist}. In addition, \dkft showed that the entire profile is also sensitive to the recent accretion rate (over one halo crossing timescale; see also \citealt{xhakaj_22} and \citealt{shin_23}). As this dependence is particularly relevant for $\rt$ and the orbiting-infalling transition region, we have binned halos by $\Gamma$ and charted the dependence of the profile parameters on this recent accretion rate. Binning by, say, concentration, would have aligned the profiles at smaller radii, around $\rs$. Given the strong correlation between $\ahalf$ and $\Gamma$, concentration-selected samples exhibit profiles similar to $\Gamma$-selected samples (Fig. 14 in \dkft).

\begin{figure}
\centering
\includegraphics[trim =  8mm 9mm 1mm 4mm, clip, scale=0.7]{\figdir/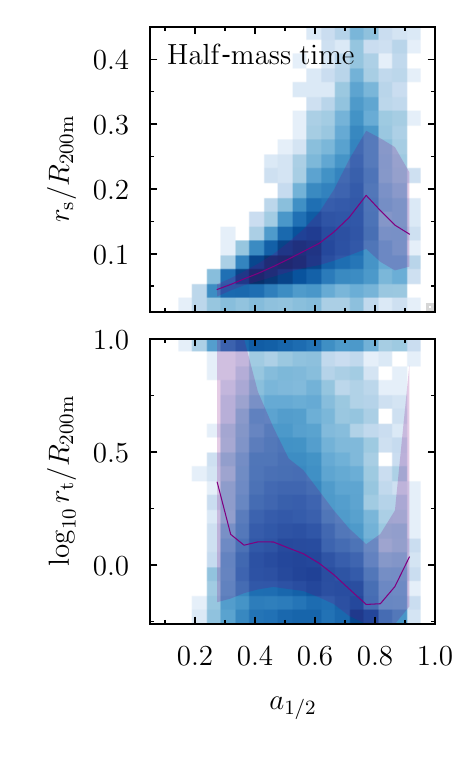}
\includegraphics[trim =  25mm 9mm 1mm 4mm, clip, scale=0.7]{\figdir/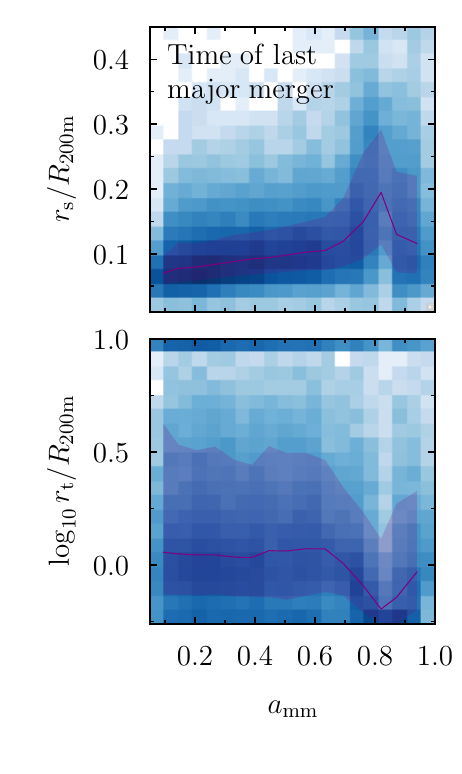}
\caption{Relation of the characteristic radii $\rs$ and $\rt$ to the formation time ($\ahalf$, left) and the time of the last major merger ($\amerge$, right). The figure shows the $\wmap$ sample at $z = 0$ only. We are using the fits to the total profile and thus not using any information from the separate orbiting profiles. We find the expected clear correlation between $\ahalf$ and $\rs$, whereas $\rt$ only weakly correlates in the expected sense that later-forming halos accrete faster today. The time of the last major merger has a much smaller influence on the profiles, but it manifests as an increased $\rs$ due to the merger dynamics and a dip in $\rt$ due to the high accretion rate at $a = 0.8$, which corresponds to half a crossing time (the average time for the merging halo to reach pericentre).}
\label{fig:timescales}
\end{figure}

To verify this picture of the history-profile connection, we now consider two timescales in a halo's history: the half-mass time $\ahalf$ and the time of the last major merger, $\amerge$ (which \consistenttrees defines as a merger with mass ratio greater than $0.3$, \citealt{behroozi_13_trees}). The latter allows us to test whether major mergers have a significant influence on the profiles, as recently suggested by the machine learning analysis of \citet{luciesmith_22_mah}. Instead of binning halo samples by $\ahalf$ and $\amerge$, Fig.~\ref{fig:timescales} shows how $\rs$ and $\rt$ relate to $\ahalf$ and $\amerge$ in individual profiles from the \wmap cosmology at $z = 0$. We use the more realistic fits to the total profile, which result in relatively reliable determinations of $\rs$ and somewhat less reliable determinations of $\rt$. The relationship between the formation time and the scale radius is easily discernible in the top-left panel, where younger halos (high $\ahalf$) exhibit larger $\rs / \rtom$ and thus smaller $\ctom$. While $\ahalf$ does correlate with $\rt$, it does so much more weakly, and in the expected way where later-forming halos have higher $\Gamma$ and thus smaller $\rt / \rtom$. 

The time of last merger, in contrast, correlates weakly with $\rs$ and $\rt$ except at very late times. Overall, we would expect that major mergers correlate with accretion in general, and that any long-term trend should thus go in the same direction as for $\ahalf$, which it does. Interestingly, both $\rs$ and $\rt$ experience significant fluctuations for $\amerge \approx 0.8$, which corresponds exactly to the time it takes particles and subhalos to cross $\rtom$ (half a halo crossing time, or $2.7\ \gyr$ at $z = 0$). We thus understand the peak in $\rs$ to be caused by the disturbance of the halo due to the major merger \citep{wang_20_concentration} and the dip in $\rt$ to correspond to the high mass accretion rate caused by the merger. For mergers that happened even more recently, the response has not yet had time to propagate to the apocenters of recently added particles, and thus to $\rt$. The parameters of the outer profile are not shown in Fig.~\ref{fig:timescales} because they do not strongly correlate with $\ahalf$ and $\amerge$. Presumably, they would be informed by the future, not past, accretion rate of halos.

These findings are highly compatible with \citet{luciesmith_22_mah}, whose machine learning algorithm identifies three features in the accretion history that strongly influence the final density profile: the formation time, recent accretion, and last major merger. The question then arises whether those three timescales should be associated with three free parameters in the density profile, i.e., whether there is a parameter beyond $\rs$ and $\rt$ that is sensitive to major mergers. Indeed, \citet{luciesmith_22_profiles} find that they need three `basis functions' to describe the profiles inside of $2\ \rtom$, which matches the three free parameters in our orbiting fit (when $\alpha$ and $\beta$ are fixed). However, it is not clear that any one basis function, or any free parameter, is clearly associated with major mergers. Given the large scatter in the relations in Fig.~\ref{fig:timescales}, the effects of mergers are probably too diverse and short-lived to be easily identified in the characteristic shapes of profiles, even though a machine learning tool can statistically detect them. 

We have established the history-profile connection on a halo-by-halo level, but cosmological analyses give broadly compatible results. For example, \citet{sanchez_22} show that the non-linear power spectra (which contain information from the infalling profiles around halos) can be accurately modelled as a function of the variance and the linear growth factor and its derivatives. This success highlights that the growth factor encodes the majority of the cosmological information probed by mildly non-linear scales. In \paperone, we interpreted the infalling profile on those scales to depend on $\nu$ (analogous to the variance $\sigma$) and $\neff$.

\subsection{On the meaning of concentration in different profile models}
\label{sec:discussion:c}

As we have discussed at length, $\rs / \rtom = 1/\ctom$ is a key property of density profiles that couples tightly to a halo's formation history. Its value, however, depends somewhat sensitively on the profile form used (NFW vs. Einasto vs. our new model) and on the fitting methodology (maximum radius, loss function, and so on). For example, NFW and Einasto concentrations differ by up to 15\% \citep[e.g.,][]{dutton_14}. We thus caution that the scale radii derived from our fits to total individual profiles are not directly analogous to those in other works. For example, \rockstar fits (out to the virial radius) result in $\rs$ being more than 50\% higher than our NFW fits on average, presumably because we fit the NFW profile out to radii that it was not designed to describe. Einasto fits (with fixed $\alpha$) broadly agree with fits of the new model, though with significant scatter. As expected, Models A and B agree almost exactly, except for very low concentrations where the difference in slope at $\rs$ (which can move slightly away from $-2$ in Model A) makes a difference. 

Given these ambiguities, it makes more sense to compare the concentrations of averaged profiles. Here, we do not detect significant differences between the Model B fits to the orbiting and total profiles, indicating that the scale radius is reliably determined by the profiles. Comparing to NFW and Einasto concentrations, we find that the new model measures up to 20\% lower $\rs$ (higher $c$) at low peak height and up to 20\% higher $\rs$ (lower $c$) at high peak height, though with some scatter in the individual datasets (see online figures). This comparison refers to NFW and Einasto fits to the total profile, i.e., with an infalling term, but out to arbitrary radii.

In summary, concentration depends moderately on the fitting function for averaged profiles, and strongly for individual profiles. We defer more detailed calibrations to future work. 

\subsection{How are the truncation and splashback radii related?}
\label{sec:discussion:rt}

In recent years, numerous works have proposed definitions of the halo boundary that are intended to be more physically motivated than spherical overdensity radii. These suggestions have included the `static mass' in shells with negative mean radial velocity \citep{cuesta_08}, the splashback radius $\rsp$ \citep[\dkft,][]{adhikari_14}, the `edge radius' where only infalling subhalos contribute to the satellite population \citep{tomooka_20, aung_21_phasespace, aung_23}, the `truncation radius' at the minimum of $r^2 \rho$ that optimizes the halo model \citep[][not to be confused with our $\rt$; see also \citealt{pizzardo_24}]{garcia_21, garcia_23}, and the `depletion radius' where the bias profile reaches its minimum \citep{fong_21}. Moreover, the splashback radius has been approximated as $\rsteep$, the radius where the logarithmic slope of the total density profile is steepest \citep{more_15}, as percentiles of the apocentre distribution of particles on their first orbit \citep{diemer_17_sparta}, and as non-spherical shells determined by local density caustics \citep{mansfield_17}. 

Fundamentally, most of these definitions try to capture the same underlying physical feature: the phase-space boundary beyond which material cannot orbit. In perfect spherical symmetry, all definitions would give the same answer because the largest apocenter radius coincides with an infinitely sharp caustic in density \citep{fillmore_84, bertschinger_85, shi_16_rsp}. In practice, all definitions encounter various difficulties: the density caustic is smoothed out and highly non-spherical, making it hard to identify in individual halo profiles and even in some stacked samples; the location of $\rsteep$ is a trade-off between the orbiting and infalling profiles; there is no one `edge' radius to which satellites can orbit, only higher and higher percentiles of the apocentre distribution; and this distribution is, at any given time, affected by the complex collapse of a halo's particles and mergers. 

The best-fit truncation radius in our profile model represents yet another measurable quantity that relates to properties of the underlying halo, such as its mass and accretion rate. In this section, we thus quantify the relationship between $\rt$, $\rsteep$, and the dynamically determined $\rsp$ as measured by \sparta.

\subsubsection{Truncation vs. steepest slope}
\label{sec:discussion:rt:rsteep}

\begin{figure}
\centering
\includegraphics[trim =  8mm 5mm 1mm 3mm, clip, scale=0.8]{\figdir/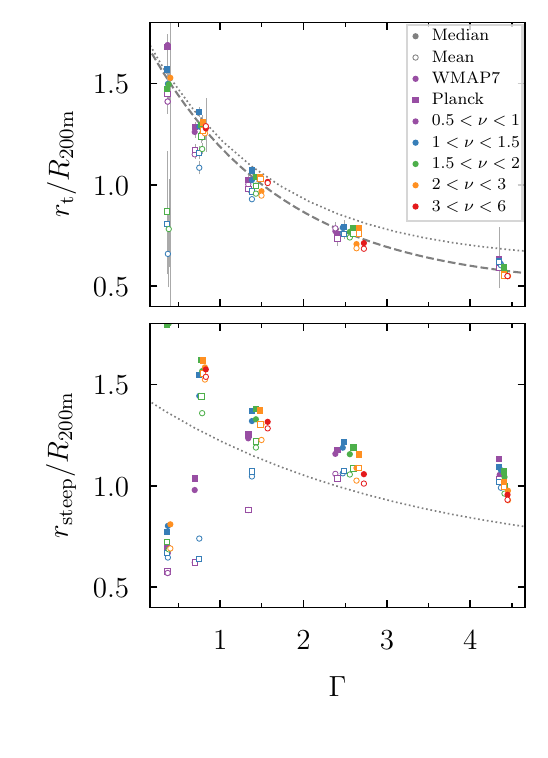}
\includegraphics[trim =  8mm 9mm 1mm 3mm, clip, scale=0.8]{\figdir/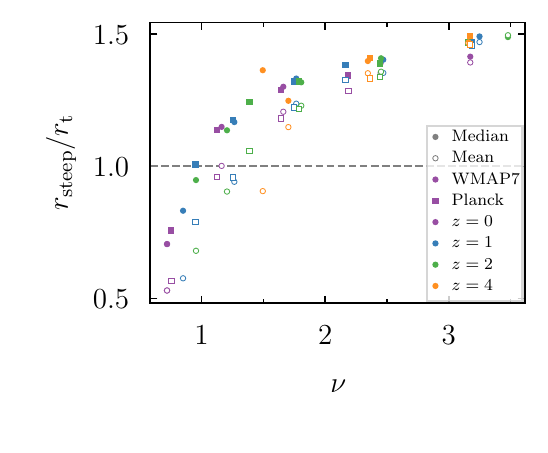}
\caption{Relation between $\rt$ and $\rsteep$, the radius where the logarithmic slope of the fitted profile is steepest. The top two panels demonstrate that $\rt$ has a cleaner relationship to the accretion rate than $\rsteep$, especially at low $\Gamma$ where $\rsteep$ is increasingly determined by the infalling profile. The dotted and dashed lines in the top panel show the relation suggested by \dkft and equation~\ref{eq:gamma_rt}, respectively. The slight disagreement arises due to the different meaning of $\rt$ in the two functions. The dotted line in the middle panel shows the relation suggested by \citealt{more_15}. Here the disagreement with our new fits arises because of the slightly different $\rsteep$ preferred by the \dkft and new fitting functions, and because the relation was not calibrated to match the lowest $\nu$ and $\Gamma$ bins. The bottom panel shows that a well-defined trend in $\rsteep / \rt$ arises not only as a function of $\Gamma$ (top panels) but also as a function of $\nu$.}
\label{fig:rsteep}
\end{figure}

In Fig.~\ref{fig:rsteep}, we test whether $\rt$ or the commonly used $\rsteep$ is more tightly connected to accretion rate. Both are determined from fits to the total profile, as they would be observationally. The top panel of Fig.~\ref{fig:rsteep} demonstrates that the $\Gamma$-$\rt$ relation is tight at all accretion rates/ Compared to Fig.~\ref{fig:rel_gamma} we have left out the self-similar Universes, which exhibit somewhat larger scatter. For the plotted \LCDM cosmologies at $z = 0$, we find 
 \begin{equation}
\label{eq:gamma_rt}
\rt / \rtom \approx 0.49 + 1.29\ e^{-\Gamma / 1.61} \,.
\end{equation}
The dotted line in Fig.~\ref{fig:rsteep} shows the relation proposed by \dkft, where we have approximately converted $\Gamma$ to the \dkft definition of accretion rate \citep{diemer_20_catalogs}. The approximate match means that the \dkft and new fitting functions assign roughly the same meaning to $\rt$, but this is only true for $\Gamma$-selected profiles in \dkft, not for their $\nu$-selected profiles (Section~\ref{sec:res_av:relations:rt}). The new function has the advantage that it can fit both types of samples with similar values of $\beta$, and thus more or less the same meaning of $\rt$.

In contrast to $\rt$, the $\rsteep$-$\Gamma$ relation breaks down at $\Gamma \lsim 2$ (middle panel of Fig.~\ref{fig:rsteep}). The reasons are two-fold: at low $\Gamma$, the edge of the orbiting term moves outwards, where it competes with a higher density of infalling material; and the second caustic due to particles on their second orbit, which resides at smaller radii, can become the location of $\rsteep$ \citep[e.g.,][]{wang_22_rsp_elucid}. We conclude that $\rt$ is the more stable indicator of accretion rate, and that one should use the best-fit $\rt$ rather than trying to determine $\rsteep$. The bottom panel of Fig.~\ref{fig:rsteep} highlights that the ratio $\rsteep / \rt$ falls on a well-defined, redshift-independent sequence with $\nu$, but the two radii can differ by up to 50\%. 

Another difficulty with $\rsteep$ is that its inferred value depends on the fitting function, or on the algorithm used to compute the slope \citep{oneil_21}. This issue is illustrated by the dotted gray line in the middle panel of Fig.~\ref{fig:rsteep}, which shows the fitting function of \citet{more_15}. This relation does not match the $\rsteep$ computed from our new fits because $\rsteep$ from \dkft fits is about 10\% lower at high $\Gamma$ and vice versa (see online figures). We find that the \dkft function is slightly better than our new model at capturing the $\rsteep$ of the actual average profiles, but the point is that $\rsteep$ is an inherently model-dependent, fickle quantity that should not be used when inferring halo properties.

\subsubsection{Truncation vs. splashback}
\label{sec:discussion:rt:rsp}

\begin{figure}
\centering
\includegraphics[trim =  9mm 20mm 2mm 1mm, clip, scale=0.62]{\figdir/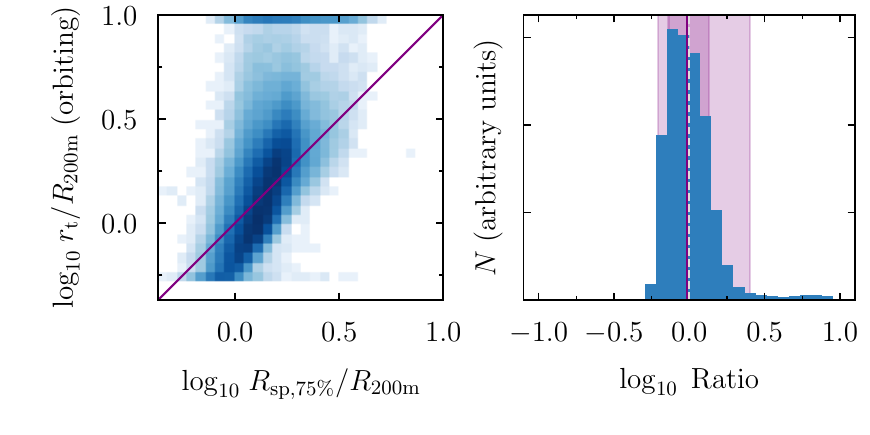}
\includegraphics[trim =  9mm 6mm 2mm 0mm, clip, scale=0.62]{\figdir/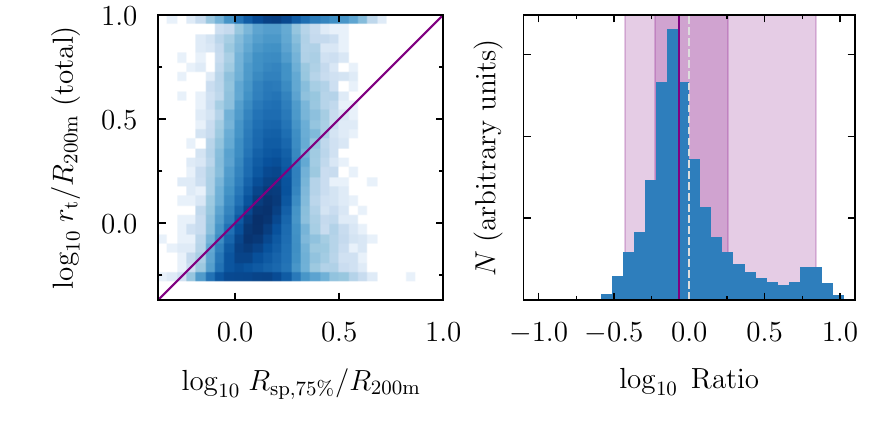}
\caption{Recovery of the splashback radius from $\rt$ as measured in fits to the separate orbiting profile (top) and the total profile (bottom). The panels have the same meaning as in Fig.~\ref{fig:ind_rs_rt}. Here, the \sparta algorithm has measured the splashback radius for each halo dynamically, defining it as the 75th percentile of the apocentres of particles on their first orbit. We have made no cuts on the sample, meaning that not all splashback radii are reliable. The underlying $\rt$ of the orbiting profile tracks the splashback radius closely, although the relationship is not exactly linear (with a $68\%$ scatter of $0.13$ dex). The more realistic $\rt$ measured in total fits exhibits much larger scatter of $0.24$ dex, confirming that it is difficult to infer $\rsp$ from individual halo profiles.}
\label{fig:rt_rsp}
\end{figure}

Part of the draw of measuring $\rsteep$ is that it is a stand-in for the splashback radius, but the relationship between $\rsteep$ and the dynamical definition of $\rsp$ is complex \citep{diemer_20_catalogs}. We might as well ask how accurately $\rt$ can approximate $\rsp$ in individual halos. As discussed in Section~\ref{sec:res_ind:total}, fits to only the orbiting term give a much more accurate estimate of $\rt$, but this success cannot be replicated in observations. Fig.~\ref{fig:rt_rsp} thus compares both values to $R_{\rm sp,75\%}$ as measured by \sparta \citep[that is, the radius that encloses 75\% of the first-orbit apocentres, see][]{diemer_17_sparta}. We find similar results for other percentile definitions, although their overall normalization shifts. Similar to Fig.~\ref{fig:ind_rs_rt}, we plot both 2D histograms of the variables and 1D histograms of their logarithmic ratio, using the entire halo sample. 

As expected, the orbiting-only $\rt$ and $\rsp$ are strongly correlated, although not exactly along the 1:1 relation. Thus, $\rt$ could be used as a proxy for $\rsp$, but both measurements demand the dynamical analysis of particle orbits (either the apocentre or pericentre counters in \sparta). When deriving $\rt$ from the entire profile, the scatter gets much larger (bottom panels of Fig.~\ref{fig:rt_rsp}). While there is still a clear correlation for the majority of haloes, there are also extreme outliers where $\rt$ simply cannot be measured. This result is not surprising, given that the inability to measure $\rsp$ from individual profiles was the motivation for algorithms such as \sparta in the first place. 

In summary, precise measurements of the splashback radius are difficult, regardless of whether it is defined as $\rsteep$ in averaged profiles or dynamically in individual halos. However, by inferring $\rt$, we can directly connect the profile fits to the accretion rate and other halo properties. We defer a comparison of $\rt$ to boundary definitions other than the splashback radius because they are either non-trivial to compute \citep[e.g.,][]{garcia_21, aung_21_phasespace} or rely on the relationship between averaged profiles and the linear matter-matter correlation function, $\xi_{\rm mm}$ \citep{fong_21}. As already shown in \dkft, we find that the infalling profiles are not well described by a bias times $\xi_{\rm mm}$ prescription, meaning that it will be difficult to physically connect the meaning of $\rt$ to estimates such as the minimum in the bias profile.


\section{Conclusions}
\label{sec:conclusion}

We have investigated the parameter space of the truncated exponential plus power-law fitting function for halo density profiles proposed in \papertwo. We have considered both averaged and individual halo profiles, as well as fits to the total profile and its separate orbiting and infalling components. Our main conclusions are as follows.

\begin{enumerate}

\item The best-fit parameters span a well-defined space where $\rt / \rtom$ and $\beta$ describe the location and sharpness of the truncation of the orbiting term. The parameters exhibit only modest degeneracies, but some of them do fall on well-defined sequences that are mostly determined by accretion rate.

\item The two characteristic scale and truncation radii, $\rs$ and $\rt$, reflect the formation time and recent accretion rate of halos, respectively. They are correlated because young halos tend to accrete faster today, but they capture separate information. Both $\rs$ and $\rt$ can be accurately inferred from stacked, total profiles. 

\item The relationship between $\alpha$ and peak height is an artefact of fitting truncated profiles with the Einasto form. When using the new fitting function, $\alpha$ is independent of $\nu$ and decreases with $\Gamma$.

\item The best-fit parameters for individual halo profiles (with $\alpha = 0.18$ and $\beta = 3$) are naturally much noisier than those based on averaged profiles, but they broadly exhibit the same inter-relations and capture the same information about the accretion rate.

\item The best-fit truncation radii of averaged profiles are tightly connected to $\Gamma$, more tightly so than the radius where the slope is steepest. Similar inferences can be made even from individual profiles, though with significant scatter.

\end{enumerate}

The key test of these conclusions will be to fit the new model to observational data using MCMC chains, both with and without some of the priors suggested by our results (such as the preferred values for $\alpha$ and $\beta$). Another important application of this work will be to build a full forward model of halo density profiles, where the profile parameters are predicted from halo properties such as their accretion history.


\section*{Acknowledgements}

I am grateful to Susmita Adhikari, Rafael Garc{\'\i}a, Edgar Salazar, Luisa Lucie-Smith, and Eduardo Rozo for productive conversations. I thank Phil Mansfield for sharing the color scheme used in several figures. This research was supported in part by the National Science Foundation under Grant numbers 2206690 and 2338388. The computations were run on the \textsc{Midway} computing cluster provided by the University of Chicago Research Computing Center and on the DeepThought2 cluster at the University of Maryland. This research extensively used the python packages \textsc{Numpy} \citep{code_numpy2}, \textsc{Scipy} \citep{code_scipy}, \textsc{Matplotlib} \citep{code_matplotlib}, and \colossus \citep{diemer_18_colossus}.


\section*{Data Availability}

The \sparta code that was used to extract the dynamically split density profiles from our simulations is publicly available in a BitBucket repository, \href{https://bitbucket.org/bdiemer/sparta}{bitbucket.org/bdiemer/sparta}. An extensive online documentation can be found at \href{https://bdiemer.bitbucket.io/sparta/}{bdiemer.bitbucket.io/sparta}. The \sparta output files (one file per simulation) are available in an hdf5 format at \href{http://erebos.astro.umd.edu/erebos/sparta}{erebos.astro.umd.edu/erebos/sparta}. A Python module to read these files is included in the \sparta code. Additional figures are provided online on the author's website at \href{http://www.benediktdiemer.com/data/}{benediktdiemer.com/data}. The full particle data for the \erebos $N$-body simulations are too large to be permanently hosted online, but they are available upon request. 


\bibliographystyle{\includedir/citestyle_mnras2}
\bibliography{\includedir/bib_mine.bib,\includedir/bib_general.bib,\includedir/bib_structure.bib,\includedir/bib_galaxies.bib,\includedir/bib_clusters.bib}

\begin{thebibliography}{97}
\expandafter\ifx\csname natexlab\endcsname\relax\def\natexlab#1{#1}\fi

\bibitem[{{Adhikari}, {Dalal} \& {Chamberlain}(2014){Adhikari}, {Dalal}, \&
  {Chamberlain}}]{adhikari_14}
{Adhikari} S., {Dalal} N., {Chamberlain} R.~T., 2014, JCAP, 11, 19

\bibitem[{{Anbajagane}, {Evrard} \& {Farahi}(2022){Anbajagane}, {Evrard}, \&
  {Farahi}}]{anbajagane_22_baryonic}
{Anbajagane} D., {Evrard} A.~E., {Farahi} A., 2022, \mnras, 509, 3441

\bibitem[{{Auger} {et~al}\mbox{.}(2013){Auger}, {Budzynski}, {Belokurov},
  {Koposov}, \& {McCarthy}}]{auger_13}
{Auger} M.~W., {Budzynski} J.~M., {Belokurov} V., {Koposov} S.~E., {McCarthy}
  I.~G., 2013, \mnras, 436, 503

\bibitem[{{Aung} {et~al}\mbox{.}(2021){Aung}, {Nagai}, {Rozo}, \&
  {Garc{\'\i}a}}]{aung_21_phasespace}
{Aung} H., {Nagai} D., {Rozo} E., {Garc{\'\i}a} R., 2021, \mnras, 502, 1041

\bibitem[{{Aung} {et~al}\mbox{.}(2023){Aung}, {Nagai}, {Rozo}, {Wolfe}, \&
  {Adhikari}}]{aung_23}
{Aung} H., {Nagai} D., {Rozo} E., {Wolfe} B., {Adhikari} S., 2023, \mnras, 521,
  3981

\bibitem[{{Baltz}, {Marshall} \& {Oguri}(2009){Baltz}, {Marshall}, \&
  {Oguri}}]{baltz_09}
{Baltz} E.~A., {Marshall} P., {Oguri} M., 2009, \jcap, 1, 15

\bibitem[{{Baxter} {et~al}\mbox{.}(2017){Baxter}, {Chang}, {Jain}, {Adhikari},
  {Dalal}, {Kravtsov}, {More}, {Rozo}, {Rykoff}, \& {Sheth}}]{baxter_17}
{Baxter} E. {et~al.}, 2017, \apj, 841, 18

\bibitem[{{Behroozi}, {Wechsler} \& {Wu}(2013){Behroozi}, {Wechsler}, \&
  {Wu}}]{behroozi_13_rockstar}
{Behroozi} P.~S., {Wechsler} R.~H., {Wu} H.-Y., 2013, \apj, 762, 109

\bibitem[{{Behroozi} {et~al}\mbox{.}(2013){Behroozi}, {Wechsler}, {Wu},
  {Busha}, {Klypin}, \& {Primack}}]{behroozi_13_trees}
{Behroozi} P.~S., {Wechsler} R.~H., {Wu} H.-Y., {Busha} M.~T., {Klypin} A.~A.,
  {Primack} J.~R., 2013, \apj, 763, 18

\bibitem[{{Bertschinger}(1985)}]{bertschinger_85}
{Bertschinger} E., 1985, \apjs, 58, 39

\bibitem[{{Brown} {et~al}\mbox{.}(2020){Brown}, {McCarthy}, {Diemer}, {Font},
  {Stafford}, \& {Pfiefer}}]{brown_20}
{Brown} S.~T., {McCarthy} I.~G., {Diemer} B., {Font} A.~S., {Stafford} S.~G.,
  {Pfiefer} S., 2020, \mnras, 495, 4994

\bibitem[{{Brown} {et~al}\mbox{.}(2022){Brown}, {McCarthy}, {Stafford}, \&
  {Font}}]{brown_22_einasto}
{Brown} S.~T., {McCarthy} I.~G., {Stafford} S.~G., {Font} A.~S., 2022, \mnras,
  509, 5685

\bibitem[{{Bullock} {et~al}\mbox{.}(2001){Bullock}, {Kolatt}, {Sigad},
  {Somerville}, {Kravtsov}, {Klypin}, {Primack}, \& {Dekel}}]{bullock_01}
{Bullock} J.~S., {Kolatt} T.~S., {Sigad} Y., {Somerville} R.~S., {Kravtsov}
  A.~V., {Klypin} A.~A., {Primack} J.~R., {Dekel} A., 2001, \mnras, 321, 559

\bibitem[{{Burkert}(1995)}]{burkert_95}
{Burkert} A., 1995, \apjl, 447, L25

\bibitem[{{Chang} {et~al}\mbox{.}(2018){Chang}, {Baxter}, {Jain},
  {S{\'a}nchez}, {Adhikari}, {Varga}, {Fang}, {Rozo}, {Rykoff}, {Kravtsov},
  {Gruen}, {Hartley}, {Huff}, {Jarvis}, {Kim}, {Prat}, {MacCrann},
  {McClintock}, {Palmese}, {Rapetti}, {Rollins}, {Samuroff}, {Sheldon},
  {Troxel}, {Wechsler}, {Zhang}, {Zuntz}, {Abbott}, {Abdalla}, {Allam},
  {Annis}, {Bechtol}, {Benoit-L{\'e}vy}, {Bernstein}, {Brooks}, {Buckley-Geer},
  {Carnero Rosell}, {Carrasco Kind}, {Carretero}, {D{\textquoteright}Andrea},
  {da Costa}, {Davis}, {Desai}, {Diehl}, {Dietrich}, {Drlica-Wagner}, {Eifler},
  {Flaugher}, {Fosalba}, {Frieman}, {Garc{\'\i}a-Bellido}, {Gaztanaga},
  {Gerdes}, {Gruendl}, {Gschwend}, {Gutierrez}, {Honscheid}, {James},
  {Jeltema}, {Krause}, {Kuehn}, {Lahav}, {Lima}, {March}, {Marshall},
  {Martini}, {Melchior}, {Menanteau}, {Miquel}, {Mohr}, {Nord}, {Ogando},
  {Plazas}, {Sanchez}, {Scarpine}, {Schindler}, {Schubnell}, {Sevilla-Noarbe},
  {Smith}, {Smith}, {Soares-Santos}, {Sobreira}, {Suchyta}, {Swanson}, {Tarle},
  {Weller}, \& {DES Collaboration}}]{chang_18}
{Chang} C. {et~al.}, 2018, \apj, 864, 83

\bibitem[{{Correa} {et~al}\mbox{.}(2015){Correa}, {Wyithe}, {Schaye}, \&
  {Duffy}}]{correa_15_b}
{Correa} C.~A., {Wyithe} J.~S.~B., {Schaye} J., {Duffy} A.~R., 2015, \mnras,
  450, 1521

\bibitem[{{Crocce}, {Pueblas} \& {Scoccimarro}(2006){Crocce}, {Pueblas}, \&
  {Scoccimarro}}]{crocce_06}
{Crocce} M., {Pueblas} S., {Scoccimarro} R., 2006, \mnras, 373, 369

\bibitem[{{Cuesta} {et~al}\mbox{.}(2008){Cuesta}, {Prada}, {Klypin}, \&
  {Moles}}]{cuesta_08}
{Cuesta} A.~J., {Prada} F., {Klypin} A., {Moles} M., 2008, \mnras, 389, 385

\bibitem[{{De Boni} {et~al}\mbox{.}(2016){De Boni}, {Serra}, {Diaferio},
  {Giocoli}, \& {Baldi}}]{deboni_16}
{De Boni} C., {Serra} A.~L., {Diaferio} A., {Giocoli} C., {Baldi} M., 2016,
  \apj, 818, 188

\bibitem[{{Dehnen}(1993)}]{dehnen_93}
{Dehnen} W., 1993, \mnras, 265, 250

\bibitem[{{Diemand} \& {Kuhlen}(2008)}]{diemand_08}
{Diemand} J., {Kuhlen} M., 2008, \apjl, 680, L25

\bibitem[{{Diemer}(2017)}]{diemer_17_sparta}
{Diemer} B., 2017, \apjs, 231, 5

\bibitem[{{Diemer}(2018)}]{diemer_18_colossus}
{Diemer} B., 2018, The Astrophysical Journal Supplement Series, 239, 35

\bibitem[{{Diemer}(2020)}]{diemer_20_catalogs}
{Diemer} B., 2020, \apjs, 251, 17

\bibitem[{{Diemer}(2022)}]{diemer_22_prof1}
{Diemer} B., 2022, \mnras, 513, 573

\bibitem[{{Diemer}(2023)}]{diemer_23_prof2}
{Diemer} B., 2023, \mnras, 519, 3292

\bibitem[{{Diemer} \& {Joyce}(2019)}]{diemer_19_cm}
{Diemer} B., {Joyce} M., 2019, \apj, 871, 168

\bibitem[{{Diemer} \& {Kravtsov}(2014)}]{diemer_14}
{Diemer} B., {Kravtsov} A.~V., 2014, \apj, 789, 1

\bibitem[{{Diemer} \& {Kravtsov}(2015)}]{diemer_15}
{Diemer} B., {Kravtsov} A.~V., 2015, \apj, 799, 108

\bibitem[{{Dutton} \& {Macci{\`o}}(2014)}]{dutton_14}
{Dutton} A.~A., {Macci{\`o}} A.~V., 2014, \mnras, 441, 3359

\bibitem[{{Eckert} {et~al}\mbox{.}(2022){Eckert}, {Ettori}, {Robertson},
  {Massey}, {Pointecouteau}, {Harvey}, \& {McCarthy}}]{eckert_22_einasto}
{Eckert} D., {Ettori} S., {Robertson} A., {Massey} R., {Pointecouteau} E.,
  {Harvey} D., {McCarthy} I.~G., 2022, \aap, 666, A41

\bibitem[{{Einasto}(1965)}]{einasto_65}
{Einasto} J., 1965, Trudy Astrofizicheskogo Instituta Alma-Ata, 5, 87

\bibitem[{{Einasto}(1969)}]{einasto_69}
{Einasto} J., 1969, Astrophysics, 5, 67

\bibitem[{{Enomoto}, {Nishimichi} \& {Taruya}(2024){Enomoto}, {Nishimichi}, \&
  {Taruya}}]{enomoto_24}
{Enomoto} Y., {Nishimichi} T., {Taruya} A., 2024, \mnras, 527, 7523

\bibitem[{{Ettori} {et~al}\mbox{.}(2010){Ettori}, {Gastaldello}, {Leccardi},
  {Molendi}, {Rossetti}, {Buote}, \& {Meneghetti}}]{ettori_10}
{Ettori} S., {Gastaldello} F., {Leccardi} A., {Molendi} S., {Rossetti} M.,
  {Buote} D., {Meneghetti} M., 2010, \aap, 524, A68

\bibitem[{{Farahi} {et~al}\mbox{.}(2018){Farahi}, {Evrard}, {McCarthy},
  {Barnes}, \& {Kay}}]{farahi_18}
{Farahi} A., {Evrard} A.~E., {McCarthy} I., {Barnes} D.~J., {Kay} S.~T., 2018,
  \mnras, 478, 2618

\bibitem[{{Fillmore} \& {Goldreich}(1984)}]{fillmore_84}
{Fillmore} J.~A., {Goldreich} P., 1984, \apj, 281, 1

\bibitem[{{Fong} \& {Han}(2021)}]{fong_21}
{Fong} M., {Han} J., 2021, \mnras, 503, 4250

\bibitem[{{Gao} {et~al}\mbox{.}(2008){Gao}, {Navarro}, {Cole}, {Frenk},
  {White}, {Springel}, {Jenkins}, \& {Neto}}]{gao_08}
{Gao} L., {Navarro} J.~F., {Cole} S., {Frenk} C.~S., {White} S.~D.~M.,
  {Springel} V., {Jenkins} A., {Neto} A.~F., 2008, \mnras, 387, 536

\bibitem[{{Garc{\'\i}a} {et~al}\mbox{.}(2021){Garc{\'\i}a}, {Rozo}, {Becker},
  \& {More}}]{garcia_21}
{Garc{\'\i}a} R., {Rozo} E., {Becker} M.~R., {More} S., 2021, \mnras, 505, 1195

\bibitem[{{Garc{\'\i}a} {et~al}\mbox{.}(2023){Garc{\'\i}a}, {Salazar}, {Rozo},
  {Adhikari}, {Aung}, {Diemer}, {Nagai}, \& {Wolfe}}]{garcia_23}
{Garc{\'\i}a} R., {Salazar} E., {Rozo} E., {Adhikari} S., {Aung} H., {Diemer}
  B., {Nagai} D., {Wolfe} B., 2023, \mnras, 521, 2464

\bibitem[{{Gunn} \& {Gott}(1972)}]{gunn_72}
{Gunn} J.~E., {Gott}, III J.~R., 1972, \apj, 176, 1

\bibitem[{{Harris} {et~al}\mbox{.}(2020){Harris}, {Millman}, {van der Walt},
  {Gommers}, {Virtanen}, {Cournapeau}, {Wieser}, {Taylor}, {Berg}, {Smith},
  {Kern}, {Picus}, {Hoyer}, {van Kerkwijk}, {Brett}, {Haldane}, {del R{\'\i}o},
  {Wiebe}, {Peterson}, {G{\'e}rard-Marchant}, {Sheppard}, {Reddy}, {Weckesser},
  {Abbasi}, {Gohlke}, \& {Oliphant}}]{code_numpy2}
{Harris} C.~R. {et~al.}, 2020, \nat, 585, 357

\bibitem[{{Hayashi} \& {White}(2008)}]{hayashi_08}
{Hayashi} E., {White} S.~D.~M., 2008, \mnras, 388, 2

\bibitem[{{Hernquist}(1990)}]{hernquist_90}
{Hernquist} L., 1990, \apj, 356, 359

\bibitem[{Hunter(2007)}]{code_matplotlib}
Hunter J.~D., 2007, Computing in Science Engineering, 9, 90

\bibitem[{{Jaffe}(1983)}]{jaffe_83}
{Jaffe} W., 1983, \mnras, 202, 995

\bibitem[{{Klypin} {et~al}\mbox{.}(2016){Klypin}, {Yepes}, {Gottl{\"o}ber},
  {Prada}, \& {He{\ss}}}]{klypin_16}
{Klypin} A., {Yepes} G., {Gottl{\"o}ber} S., {Prada} F., {He{\ss}} S., 2016,
  \mnras, 457, 4340

\bibitem[{{Klypin}, {Trujillo-Gomez} \& {Primack}(2011){Klypin},
  {Trujillo-Gomez}, \& {Primack}}]{klypin_11}
{Klypin} A.~A., {Trujillo-Gomez} S., {Primack} J., 2011, \apj, 740, 102

\bibitem[{{Knollmann}, {Power} \& {Knebe}(2008){Knollmann}, {Power}, \&
  {Knebe}}]{knollmann_08}
{Knollmann} S.~R., {Power} C., {Knebe} A., 2008, \mnras, 385, 545

\bibitem[{{Komatsu} {et~al}\mbox{.}(2011){Komatsu}, {Smith}, {Dunkley},
  {Bennett}, {Gold}, {Hinshaw}, {Jarosik}, {Larson}, {Nolta}, {Page},
  {Spergel}, {Halpern}, {Hill}, {Kogut}, {Limon}, {Meyer}, {Odegard}, {Tucker},
  {Weiland}, {Wollack}, \& {Wright}}]{komatsu_11}
{Komatsu} E. {et~al.}, 2011, \apjs, 192, 18

\bibitem[{{Lewis}, {Challinor} \& {Lasenby}(2000){Lewis}, {Challinor}, \&
  {Lasenby}}]{lewis_00}
{Lewis} A., {Challinor} A., {Lasenby} A., 2000, \apj, 538, 473

\bibitem[{{L{\'o}pez-Cano} {et~al}\mbox{.}(2022){L{\'o}pez-Cano}, {Angulo},
  {Ludlow}, {Zennaro}, {Contreras}, {Chaves-Montero}, \&
  {Aric{\`o}}}]{lopezcano_22}
{L{\'o}pez-Cano} D., {Angulo} R.~E., {Ludlow} A.~D., {Zennaro} M., {Contreras}
  S., {Chaves-Montero} J., {Aric{\`o}} G., 2022, \mnras, 517, 2000

\bibitem[{{Lucie-Smith}, {Adhikari} \& {Wechsler}(2022){Lucie-Smith},
  {Adhikari}, \& {Wechsler}}]{luciesmith_22_mah}
{Lucie-Smith} L., {Adhikari} S., {Wechsler} R.~H., 2022, \mnras, 515, 2164

\bibitem[{{Lucie-Smith}, {Peiris} \& {Pontzen}(2024){Lucie-Smith}, {Peiris}, \&
  {Pontzen}}]{luciesmith_24}
{Lucie-Smith} L., {Peiris} H.~V., {Pontzen} A., 2024, \prl, 132, 031001

\bibitem[{{Lucie-Smith} {et~al}\mbox{.}(2022){Lucie-Smith}, {Peiris},
  {Pontzen}, {Nord}, {Thiyagalingam}, \& {Piras}}]{luciesmith_22_profiles}
{Lucie-Smith} L., {Peiris} H.~V., {Pontzen} A., {Nord} B., {Thiyagalingam} J.,
  {Piras} D., 2022, \prd, 105, 103533

\bibitem[{{Ludlow} \& {Angulo}(2017)}]{ludlow_17}
{Ludlow} A.~D., {Angulo} R.~E., 2017, \mnras, 465, L84

\bibitem[{{Ludlow} {et~al}\mbox{.}(2013){Ludlow}, {Navarro}, {Boylan-Kolchin},
  {Bett}, {Angulo}, {Li}, {White}, {Frenk}, \& {Springel}}]{ludlow_13}
{Ludlow} A.~D. {et~al.}, 2013, \mnras, 432, 1103

\bibitem[{{Ludlow}, {Schaye} \& {Bower}(2019){Ludlow}, {Schaye}, \&
  {Bower}}]{ludlow_19}
{Ludlow} A.~D., {Schaye} J., {Bower} R., 2019, \mnras, 488, 3663

\bibitem[{{Mansfield} \& {Avestruz}(2021)}]{mansfield_21_resolution}
{Mansfield} P., {Avestruz} C., 2021, \mnras, 500, 3309

\bibitem[{{Mansfield}, {Kravtsov} \& {Diemer}(2017){Mansfield}, {Kravtsov}, \&
  {Diemer}}]{mansfield_17}
{Mansfield} P., {Kravtsov} A.~V., {Diemer} B., 2017, \apj, 841, 34

\bibitem[{{Merritt} {et~al}\mbox{.}(2006){Merritt}, {Graham}, {Moore},
  {Diemand}, \& {Terzi{\'c}}}]{merritt_06}
{Merritt} D., {Graham} A.~W., {Moore} B., {Diemand} J., {Terzi{\'c}} B., 2006,
  \aj, 132, 2685

\bibitem[{{More}, {Diemer} \& {Kravtsov}(2015){More}, {Diemer}, \&
  {Kravtsov}}]{more_15}
{More} S., {Diemer} B., {Kravtsov} A.~V., 2015, \apj, 810, 36

\bibitem[{{More} {et~al}\mbox{.}(2016){More}, {Miyatake}, {Takada}, {Diemer},
  {Kravtsov}, {Dalal}, {More}, {Murata}, {Mandelbaum}, {Rozo}, {Rykoff},
  {Oguri}, \& {Spergel}}]{more_16}
{More} S. {et~al.}, 2016, \apj, 825, 39

\bibitem[{{Muni} {et~al}\mbox{.}(2024){Muni}, {Pontzen}, {Sanders}, {Rey},
  {Read}, \& {Agertz}}]{muni_24}
{Muni} C., {Pontzen} A., {Sanders} J.~L., {Rey} M.~P., {Read} J.~I., {Agertz}
  O., 2024, \mnras, 527, 9250

\bibitem[{{Navarro}, {Frenk} \& {White}(1995){Navarro}, {Frenk}, \&
  {White}}]{navarro_95}
{Navarro} J.~F., {Frenk} C.~S., {White} S.~D.~M., 1995, \mnras, 275, 720

\bibitem[{{Navarro}, {Frenk} \& {White}(1996){Navarro}, {Frenk}, \&
  {White}}]{navarro_96}
{Navarro} J.~F., {Frenk} C.~S., {White} S.~D.~M., 1996, \apj, 462, 563

\bibitem[{{Navarro}, {Frenk} \& {White}(1997){Navarro}, {Frenk}, \&
  {White}}]{navarro_97}
{Navarro} J.~F., {Frenk} C.~S., {White} S.~D.~M., 1997, \apj, 490, 493

\bibitem[{{Navarro} {et~al}\mbox{.}(2010){Navarro}, {Ludlow}, {Springel},
  {Wang}, {Vogelsberger}, {White}, {Jenkins}, {Frenk}, \& {Helmi}}]{navarro_10}
{Navarro} J.~F. {et~al.}, 2010, \mnras, 402, 21

\bibitem[{{Nipoti}(2015)}]{nipoti_15}
{Nipoti} C., 2015, \apjl, 805, L16

\bibitem[{{Oguri} {et~al}\mbox{.}(2012){Oguri}, {Bayliss}, {Dahle}, {Sharon},
  {Gladders}, {Natarajan}, {Hennawi}, \& {Koester}}]{oguri_12}
{Oguri} M., {Bayliss} M.~B., {Dahle} H., {Sharon} K., {Gladders} M.~D.,
  {Natarajan} P., {Hennawi} J.~F., {Koester} B.~P., 2012, \mnras, 420, 3213

\bibitem[{{O'Neil} {et~al}\mbox{.}(2021){O'Neil}, {Barnes}, {Vogelsberger}, \&
  {Diemer}}]{oneil_21}
{O'Neil} S., {Barnes} D.~J., {Vogelsberger} M., {Diemer} B., 2021, \mnras, 504,
  4649

\bibitem[{{Pizzardo}, {Diaferio} \& {Rines}(2024){Pizzardo}, {Diaferio}, \&
  {Rines}}]{pizzardo_24}
{Pizzardo} M., {Diaferio} A., {Rines} K.~J., 2024, \aap, 682, A80

\bibitem[{{Planck Collaboration} {et~al}\mbox{.}(2014){Planck Collaboration},
  {Ade}, {Aghanim}, {Armitage-Caplan}, {Arnaud}, {Ashdown}, {Atrio-Barandela},
  {Aumont}, {Baccigalupi}, {Banday}, \& et~al.}]{planck_14}
{Planck Collaboration} {et~al.}, 2014, \aap, 571, A16

\bibitem[{{Prada} {et~al}\mbox{.}(2012){Prada}, {Klypin}, {Cuesta},
  {Betancort-Rijo}, \& {Primack}}]{prada_12}
{Prada} F., {Klypin} A.~A., {Cuesta} A.~J., {Betancort-Rijo} J.~E., {Primack}
  J., 2012, \mnras, 423, 3018

\bibitem[{{Salazar} {et~al}\mbox{.}(2024){Salazar}, {Rozo}, {Garc{\'\i}a},
  {Kokron}, {Adhikari}, {Diemer}, \& {Osinga}}]{salazar_24}
{Salazar} E.~M., {Rozo} E., {Garc{\'\i}a} R., {Kokron} N., {Adhikari} S.,
  {Diemer} B., {Osinga} C., 2024, arXiv e-prints, arXiv:2406.04054

\bibitem[{{S{\'a}nchez} {et~al}\mbox{.}(2022){S{\'a}nchez}, {Ruiz}, {Jara}, \&
  {Padilla}}]{sanchez_22}
{S{\'a}nchez} A.~G., {Ruiz} A.~N., {Jara} J.~G., {Padilla} N.~D., 2022, \mnras,
  514, 5673

\bibitem[{{Shi}(2016)}]{shi_16_rsp}
{Shi} X., 2016, \mnras, 459, 3711

\bibitem[{{Shin} {et~al}\mbox{.}(2019){Shin}, {Adhikari}, {Baxter}, {Chang},
  {Jain}, {Battaglia}, {Bleem}, {Bocquet}, {DeRose}, {Gruen}, {Hilton},
  {Kravtsov}, {McClintock}, {Rozo}, {Rykoff}, {Varga}, {Wechsler}, {Wu},
  {Zhang}, {Aiola}, {Allam}, {Bechtol}, {Benson}, {Bertin}, {Bond}, {Brodwin},
  {Brooks}, {Buckley-Geer}, {Burke}, {Carlstrom}, {Carnero Rosell}, {Carrasco
  Kind}, {Carretero}, {Castander}, {Choi}, {Cunha}, {Crawford}, {da Costa}, {De
  Vicente}, {Desai}, {Devlin}, {Dietrich}, {Doel}, {Dunkley}, {Eifler},
  {Evrard}, {Flaugher}, {Fosalba}, {Gallardo}, {Garc{\'\i}a-Bellido},
  {Gaztanaga}, {Gerdes}, {Gralla}, {Gruendl}, {Gschwend}, {Gupta}, {Gutierrez},
  {Hartley}, {Hill}, {Ho}, {Hollowood}, {Honscheid}, {Hoyle}, {Huffenberger},
  {Hughes}, {James}, {Jeltema}, {Kim}, {Krause}, {Kuehn}, {Lahav}, {Lima},
  {Madhavacheril}, {Maia}, {Marshall}, {Maurin}, {McMahon}, {Menanteau},
  {Miller}, {Miquel}, {Mohr}, {Naess}, {Nati}, {Newburgh}, {Niemack}, {Ogando},
  {Page}, {Partridge}, {Patil}, {Plazas}, {Rapetti}, {Reichardt}, {Romer},
  {Sanchez}, {Scarpine}, {Schindler}, {Serrano}, {Smith}, {Smith},
  {Soares-Santos}, {Sobreira}, {Staggs}, {Stark}, {Stein}, {Suchyta},
  {Swanson}, {Tarle}, {Thomas}, {van Engelen}, {Wollack}, \&
  {Xu}}]{shin_19_rsp}
{Shin} T. {et~al.}, 2019, \mnras, 487, 2900

\bibitem[{{Shin} {et~al}\mbox{.}(2021){Shin}, {Jain}, {Adhikari}, {Baxter},
  {Chang}, {Pandey}, {Salcedo}, {Weinberg}, {Amsellem}, {Battaglia},
  {Belyakov}, {Dacunha}, {Goldstein}, {Kravtsov}, {Varga}, {Abbott}, {Aguena},
  {Alarcon}, {Allam}, {Amon}, {Andrade-Oliveira}, {Annis}, {Bacon}, {Bechtol},
  {Becker}, {Bernstein}, {Bertin}, {Bocquet}, {Bond}, {Brooks}, {Buckley-Geer},
  {Burke}, {Campos}, {Rosell}, {Kind}, {Carretero}, {Chen}, {Choi}, {Costanzi},
  {da Costa}, {DeRose}, {Desai}, {De Vicente}, {Devlin}, {Diehl}, {Dietrich},
  {Dodelson}, {Doel}, {Doux}, {Drlica-Wagner}, {Eckert}, {Elvin-Poole},
  {Everett}, {Ferraro}, {Ferrero}, {Fert{\'e}}, {Flaugher}, {Frieman},
  {Gallardo}, {Gatti}, {Gaztanaga}, {Gerdes}, {Gruen}, {Gruendl}, {Gutierrez},
  {Harrison}, {Hartley}, {Hill}, {Hilton}, {Hinton}, {Hollowood}, {Hughes},
  {James}, {Jarvis}, {Jeltema}, {Koopman}, {Krause}, {Kuehn}, {Kuropatkin},
  {Lahav}, {Lima}, {Lokken}, {MacCrann}, {Madhavacheril}, {Maia}, {McCullough},
  {McMahon}, {Melchior}, {Menanteau}, {Miquel}, {Mohr}, {Moodley}, {Morgan},
  {Myles}, {Nati}, {Navarro-Alsina}, {Niemack}, {Ogando}, {Page}, {Palmese},
  {Partridge}, {Paz-Chinch{\'o}n}, {Pereira}, {Pieres}, {Malag{\'o}n}, {Prat},
  {Raveri}, {Rodriguez-Monroy}, {Rollins}, {Romer}, {Rykoff}, {Salatino},
  {S{\'a}nchez}, {Sanchez}, {Santiago}, {Scarpine}, {Schillaci}, {Secco},
  {Serrano}, {Sevilla-Noarbe}, {Sheldon}, {Sherwin}, {Sif{\'o}n}, {Smith},
  {Soares-Santos}, {Staggs}, {Suchyta}, {Swanson}, {Tarle}, {Thomas}, {To},
  {Troxel}, {Tutusaus}, {Vavagiakis}, {Weller}, {Wollack}, {Yanny}, {Yin}, \&
  {Zhang}}]{shin_21}
{Shin} T. {et~al.}, 2021, \mnras, 507, 5758

\bibitem[{{Shin} \& {Diemer}(2023)}]{shin_23}
{Shin} T.-h., {Diemer} B., 2023, \mnras, 521, 5570

\bibitem[{{Springel}(2005)}]{springel_05_gadget2}
{Springel} V., 2005, \mnras, 364, 1105

\bibitem[{{Sugiura} {et~al}\mbox{.}(2020){Sugiura}, {Nishimichi}, {Rasera}, \&
  {Taruya}}]{sugiura_20}
{Sugiura} H., {Nishimichi} T., {Rasera} Y., {Taruya} A., 2020, \mnras, 493,
  2765

\bibitem[{{Tasitsiomi} {et~al}\mbox{.}(2004){Tasitsiomi}, {Kravtsov},
  {Gottl{\"o}ber}, \& {Klypin}}]{tasitsiomi_04_clusterprof}
{Tasitsiomi} A., {Kravtsov} A.~V., {Gottl{\"o}ber} S., {Klypin} A.~A., 2004,
  \apj, 607, 125

\bibitem[{{Tomooka} {et~al}\mbox{.}(2020){Tomooka}, {Rozo}, {Wagoner}, {Aung},
  {Nagai}, \& {Safonova}}]{tomooka_20}
{Tomooka} P., {Rozo} E., {Wagoner} E.~L., {Aung} H., {Nagai} D., {Safonova} S.,
  2020, \mnras, 499, 1291

\bibitem[{{Udrescu} {et~al}\mbox{.}(2019){Udrescu}, {Dutton}, {Macci{\`o}}, \&
  {Buck}}]{udrescu_19}
{Udrescu} S.~M., {Dutton} A.~A., {Macci{\`o}} A.~V., {Buck} T., 2019, \mnras,
  482, 5259

\bibitem[{{Umetsu}(2020)}]{umetsu_20_review}
{Umetsu} K., 2020, \aapr, 28, 7

\bibitem[{{Umetsu} \& {Diemer}(2017)}]{umetsu_17}
{Umetsu} K., {Diemer} B., 2017, \apj, 836, 231

\bibitem[{{Virtanen} {et~al}\mbox{.}(2020){Virtanen}, {Gommers}, {Oliphant},
  {Haberland}, {Reddy}, {Cournapeau}, {Burovski}, {Peterson}, {Weckesser},
  {Bright}, {van der Walt}, {Brett}, {Wilson}, {Millman}, {Mayorov}, {Nelson},
  {Jones}, {Kern}, {Larson}, {Carey}, {Polat}, {Feng}, {Moore}, {VanderPlas},
  {Laxalde}, {Perktold}, {Cimrman}, {Henriksen}, {Quintero}, {Harris},
  {Archibald}, {Ribeiro}, {Pedregosa}, {van Mulbregt}, \& {SciPy 1. 0
  Contributors}}]{code_scipy}
{Virtanen} P. {et~al.}, 2020, Nature Methods, 17, 261

\bibitem[{{Wang} {et~al}\mbox{.}(2020{\natexlab{a}}){Wang}, {Bose}, {Frenk},
  {Gao}, {Jenkins}, {Springel}, \& {White}}]{wang_20_zoom}
{Wang} J., {Bose} S., {Frenk} C.~S., {Gao} L., {Jenkins} A., {Springel} V.,
  {White} S.~D.~M., 2020{\natexlab{a}}, \nat, 585, 39

\bibitem[{{Wang} {et~al}\mbox{.}(2020{\natexlab{b}}){Wang}, {Mao}, {Zentner},
  {Lange}, {van den Bosch}, \& {Wechsler}}]{wang_20_concentration}
{Wang} K., {Mao} Y.-Y., {Zentner} A.~R., {Lange} J.~U., {van den Bosch} F.~C.,
  {Wechsler} R.~H., 2020{\natexlab{b}}, \mnras, 498, 4450

\bibitem[{{Wang}, {Wang} \& {Mo}(2022){Wang}, {Wang}, \&
  {Mo}}]{wang_22_rsp_elucid}
{Wang} X., {Wang} H., {Mo} H.~J., 2022, \aap, 667, A99

\bibitem[{{Wechsler} {et~al}\mbox{.}(2002){Wechsler}, {Bullock}, {Primack},
  {Kravtsov}, \& {Dekel}}]{wechsler_02}
{Wechsler} R.~H., {Bullock} J.~S., {Primack} J.~R., {Kravtsov} A.~V., {Dekel}
  A., 2002, \apj, 568, 52

\bibitem[{{Wiesner} {et~al}\mbox{.}(2012){Wiesner}, {Lin}, {Allam}, {Annis},
  {Buckley-Geer}, {Diehl}, {Kubik}, {Kubo}, \& {Tucker}}]{wiesner_12}
{Wiesner} M.~P. {et~al.}, 2012, \apj, 761, 1

\bibitem[{{Xhakaj} {et~al}\mbox{.}(2022){Xhakaj}, {Leauthaud}, {Lange},
  {Hearin}, {Diemer}, \& {Dalal}}]{xhakaj_22}
{Xhakaj} E., {Leauthaud} A., {Lange} J., {Hearin} A., {Diemer} B., {Dalal} N.,
  2022, \mnras, 514, 2876

\bibitem[{{Zhao} {et~al}\mbox{.}(2009){Zhao}, {Jing}, {Mo}, \&
  {B{\"o}rner}}]{zhao_09_acchist}
{Zhao} D.~H., {Jing} Y.~P., {Mo} H.~J., {B{\"o}rner} G., 2009, \apj, 707, 354

\bibitem[{{Zhou} \& {Han}(2024)}]{zhou_24}
{Zhou} Y., {Han} J., 2024, arXiv e-prints, arXiv:2407.08381

\end{thebibliography}




\bsp
\label{lastpage}
\end{document}